\documentclass{article}
\usepackage{graphicx}
\usepackage{epsf}
\usepackage{epsfig}
\usepackage{amssymb}
\usepackage{amsmath}
\makeatletter
\@addtoreset{equation}{section}
\makeatother

\def\bea{\begin{eqnarray}}
\def\eea{\end{eqnarray}}
\def\be{\begin{equation}}
\def\ee{\end{equation}}

\def\wgta#1#2#3#4{\hbox{\rlap{\lower.35cm\hbox{$#1$}}
\hskip.2cm\rlap{\raise.25cm\hbox{$#2$}}
\rlap{\vrule width1.3cm height.4pt}
\hskip.55cm\rlap{\lower.6cm\hbox{\vrule width.4pt height1.2cm}}
\hskip.15cm
\rlap{\raise.25cm\hbox{$#3$}}\hskip.25cm\lower.35cm\hbox{$#4$}\hskip.6cm}}

\def\wgtb#1#2#3#4{\hbox{\rlap{\raise.25cm\hbox{$#2$}}
\hskip.2cm\rlap{\lower.35cm\hbox{$#1$}}
\rlap{\vrule width1.3cm height.4pt}
\hskip.55cm\rlap{\lower.6cm\hbox{\vrule width.4pt height1.2cm}}
\hskip.15cm
\rlap{\lower.35cm\hbox{$#4$}}\hskip.25cm\raise.25cm\hbox{$#3$}\hskip.6cm}}

\def\begeqar{\begin{eqnarray}}
\def\endeqar{\end{eqnarray}}
\title{Conformal boundary loop models}
\author{Jesper Lykke Jacobsen${}^{1,2}$ and
        Hubert Saleur${}^{2,3}$ \\[2.0mm]
${}^1$ LPTMS, Universit\'e Paris-Sud, B\^atiment 100, \\
Orsay, 91405, France \\
${}^2$ Service de Physique Th\'eorique, CEA Saclay, \\
Gif Sur Yvette, 91191, France \\
${}^3$ Department of Physics and Astronomy,
University of Southern California, \\
Los Angeles, CA 90089, USA}

\begin{document}

\maketitle

\begin{abstract}

  We study a model of densely packed self-avoiding loops on the
  annulus, related to the Temperley Lieb algebra with an extra
  idempotent boundary generator.  Four different weights are given to
  the loops, depending on their homotopy class and whether they touch
  the outer rim of the annulus. When the weight of a contractible bulk
  loop $x \equiv q + q^{-1} \in (-2,2]$, this model is conformally
  invariant for any real weight of the remaining three
  parameters.  We classify the conformal boundary conditions and give
  exact expressions for the corresponding boundary scaling dimensions.
  The amplitudes with which the sectors with any prescribed number and
  types of non contractible loops appear in the full partition
  function $Z$ are computed rigorously. Based on this, we write a
  number of identities involving $Z$ which hold true for any finite
  size. When the weight of a contractible boundary loop $y$ takes
  certain discrete values, $y_r \equiv \frac{[r+1]_q}{[r]_q}$ with $r$ integer, other
  identities involving the standard characters $K_{r,s}$ of the 
  Virasoro algebra are established.  The
  connection with Dirichlet and Neumann boundary conditions in the
  $O(n)$ model is discussed in detail, and new scaling dimensions are
  derived. When $q$ is a root of unity and $y=y_r$, exact connections
  with the $A_m$ type RSOS model are made. These involve precise
  relations between the spectra of the loop and RSOS model transfer
  matrices, valid in finite size. Finally, the results where $y=y_r$
  are related to the theory of Temperley Lieb cabling.

\end{abstract}

\bigskip

\noindent SPhT-T06/155

\smallskip

\section{Introduction}

Boundary conformal field theories (CFT) have lately  played an increasingly 
important role in statistical mechanics, condensed matter physics and 
string theory. In statistical mechanics, they appear in
most probabilistic applications of geometrical models (see, e.g., 
\cite{Cardy06} for a recent example), 
in particular through SLE \cite{BauerBernard}. In condensed matter, 
they contain all the information about fixed points in theories which 
are gapless in the bulk, such as Kondo systems or edge states in the 
fractional quantum Hall effect (see \cite{SaleurHouches} for a 
review). In string theory, they provide for instance microscopic techniques 
to study D-branes in curved backgrounds \cite{VolkerReview1}. On top of this, the 
study of boundary aspects is a crucial component of understanding and 
classifying CFTs at large \cite{JBZ}, and has lately  played a crucial role in the 
solution of non rational CFTs \cite{VolkerReview2}.

While progress in understanding conformal boundary conditions for 
rational  CFTs has been considerable, the 
situation is not so satisfactory for non rational theories, which are 
however all too frequent in statistical mechanics applications. A 
case in point concerns the loop or cluster models, which---in one guise 
or another---are hidden behind most simple models of interest, such as
$Q$-state Potts models, $O(n)$ models, RSOS models, polymers,  and
percolation. We are not aware of answers to most questions one 
might ask in this context, such as ``what are all the conformal 
invariant boundary conditions'', or ``what are the partition 
functions for the $O(n)$ models with Neumann boundary conditions'', 
etc. The origin of this difficulty lies in our lack of understanding 
of the bulk CFTs, which exhibit non-rational, logarithmic features, 
and for which too little is known. Bulk exponents turned out to be 
tractable thanks to the Coulomb gas technique, but this technique has 
not been generalized to the boundary case with sufficient control yet 
(see \cite{Cardy06} for recent progress in this direction).

We put forward in this paper a proposal for what we believe are all 
the conformal boundary conditions of dense loop models. For each 
of those we determine the critical exponents, operator content and 
boundary partition functions, some of which have interesting 
probabilistic or geometrical interpretations. 

There seems to be much substance behind the results we uncover, and 
we hope to get back to the question in more details in the near 
future. In the present paper, we only present the leading arguments---which
are based on previously published but unexploited results, as well 
as algebraic considerations---together with intensive numerical 
checks and some combinatorial proofs. 

To help the reader, we now give a quick summary of our results and
notations.  The {\em boundary loop model} (BLM) to be studied is
defined on a tilted square lattice (see Fig.~\ref{vertices}), wrapped
on an annulus of width $N$ strands and circumference $M$ lattice
spacings. Loops  cover all the edges, and 
interact in a specific way with the outer rim of the
annulus, whereas they are simply reflected by the inner rim (free
boundary conditions). We denote by $L$ the number of non contractible
loops (note that $L$ and $N$ must have the same parity). Any loop has
one of four weights ($x$, $y$, $l$ or $m$, see Fig.~\ref{partfunct}):
$l$ (resp.\ $m$) for a non contractible loop never touching (resp.\
touching at least once) the outer rim, and similarly $x$ (resp.\ $y$)
for contractible loops. We parametrize $x=q+q^{-1} \in (-2,2]$ by
$q={\rm e}^{i\pi/(p+1)}$ ($p$ real); the model is then critical with central
charge (\ref{central}) for any real values of $y$, $l$, $m$ and is
endowed with the $U_q(sl_2)$ quantum group symmetry.  We further
parametrize $y=y(r)$ as in (\ref{yloop}). Our central claim is that
for any {\em real} $r$, and any $L$, there are two (distinct for $L>0$) conformal
boundary conditions: {\em blobbed} (resp.\ {\em unblobbed}) in which
the outermost non contractible loop is required to (resp.\ required
not to) touch the outer rim of the annulus. (When $L=0$ the two cases
coincide.) The spectrum generating functions in these two cases are
(\ref{specgenfunc}), and the boundary conformal weights (critical
exponents) $h_{r,r\pm L}$ are read off from (\ref{Kac}). They combine
to form the BLM partition function $Z$ through the amplitudes
(\ref{guess}). When $p \ge 1$ is integer,  and when further $r=1,2,\ldots,p$
the BLM model  can be
related to an RSOS model of the $A_p$ type with specific boundary
conditions (three columns of fixed heights, see Fig.~\ref{RSOSconf}) through the rules
(\ref{RSOSrules}). In the latter case, $Z$ can be written as a sum
(\ref{niceZ}) over irreducible representations of the Virasoro
algebra.

The paper is organized as follows. In Section~\ref{sec:blob} we review
the algebraic framework used in our study (the blob algebra) along
with a few key results.  In Section~\ref{sec:BLM} we define the BLM,
classify its conformal boundary conditions, and give exact results for
the associated critical exponents. Two appendices present a rigorous
result on the amplitudes of the transfer matrix eigenvalues.  This is
used in Section~\ref{sec:Z} to write a number of exact
identities---exact in finite size---relating $Z$ to Virasoro
characters $K_{r,s}$.  In Section~\ref{sec:Neumann} we discuss the
case of Neumann boundary conditions for the loop model, identifying in
particular the Neumann to Dirichlet boundary condition changing
field. The relations to RSOS models are discussed in
Section~\ref{sec:RSOS}. Finally, in Section~\ref{sec:cabling}, we
comment on the relation between the blob algebra and the theory of
Temperley Lieb cabling. Our conclusions---and the prospects for (much)
further work---are given in Section~\ref{sec:conclusions}.

\section{The blob algebra}
\label{sec:blob}

 The Temperley Lieb (TL) algebra $\mathcal{T}_{N}(x)$ on $N$ strands is defined by the
generators $e_i$ ($i=1,2,\ldots,N-1$) acting on strands $i$ and $i+1$
and satisfying the well-known relations
\begin{eqnarray}
 e_i e_j &=& e_j e_i \mbox{ for $|i-j| \ge 2$} \nonumber \\
 e_i e_{i \pm 1} e_i &=& e_i \nonumber \\
 e_i^2 &=& x e_i
\label{TLalg}
\end{eqnarray}
In \cite{MartinSaleur} this was generalized into the
two parameter ``blob algebra'' $\mathcal{B}_{N}(x,y)$, having an extra generator
$b$, and satisfying in addition the relations
 \begin{eqnarray}
 b^{2}&=&b\nonumber\\
 e_{1} b e_{1}&=& y e_{1}\nonumber\\
 e_i b &=& b e_i~~~ \mbox{ for $i=2,3,\ldots,N-1$}
\label{blobalg}
\end{eqnarray}

 \begin{figure}
\begin{center}
  \leavevmode
  \includegraphics[scale=.45,angle=0 ]{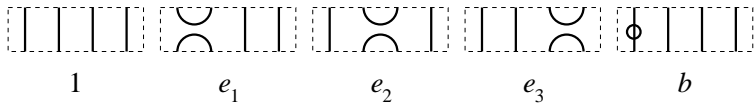}
\end{center}
\protect\caption{Graphical representation of the action of the generators of
  the blob algebra, here shown for a system of $N=4$ strands.}
\label{generators}
\end{figure}

\begin{figure}
\begin{center}
  \leavevmode
  \includegraphics[scale=.45,angle=0 ]{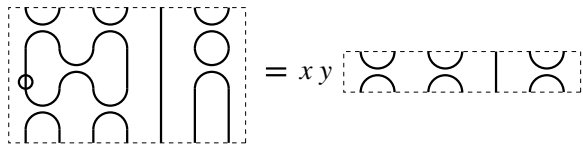}
\end{center}
\protect\caption{The word $e_1 e_3 e_6 e_2 e_6 b e_1 e_3 = x y 
  e_1 e_3 e_6$ in the blob algebra $\mathcal{B}(x,y)$ on $N=7$ strands.}
\label{blob}
\end{figure}

For $\mathcal{T}_{N}(x)$ it is well-known how to interpret these algebraic relations
graphically in terms of the strands. The extra generator $b$ marks the
leftmost strand by adding a ``blob'' to it (see Figure \ref{generators}). In
this graphical representation any completed loop may be taken out and replaced
by its corresponding weight (see Figure \ref{blob}). A
loop with no blob gets the usual weight of $x$, while a
loop with a blob gets a modified weight $y$. Note that several blobs on
the {\em same} loop reduce to a single blob. Obviously, only
loops touching the left border can be blobbed.
 
 This algebra has given rise to much work in recent years in the 
 mathematical literature \cite{MathBlob}. It has also 
 been studied in the context of boundary conformal field theory 
 \cite{Nicholsmult} with results that have some small overlap with ours. 
 (In \cite{Nicholsmult} this algebra is called the ``one boundary TL 
 algebra'', a 
 name we shall not adopt.)
 The blob algebra 
 is in fact a quotient of the more general affine Hecke algebra (like 
 TL itself is a quotient of the ordinary Hecke algebra). 
 
 The representation theory of the two parameter algebra is richer than 
 the representation theory of the TL algebra. For a given 
 $x=q+q^{-1}$, and assuming first that $q=e^{i\gamma}$ is not a root of unity, 
 exceptional cases occur whenever 
 \begin{equation}
     y={\sin(r\pm 1)\gamma\over \sin r\gamma}, \quad \mbox{$r$ integer}
\label{yloopsigned} 
\end{equation}
[In the original paper \cite{MartinSaleur}
this corresponds to $\eta=\mp r\gamma\hbox{ mod }\pi$
in the basic equation
$y=\left( q-q^{-1}e^{2i\eta} \right) / \left(1-e^{2i\eta} \right)$.]
   
In the case
\begin{equation}
 y={\sin(r+ 1)\gamma\over \sin r\gamma}, \quad \mbox{$r$ integer}
\label{yloop}
\end{equation}
the spectrum of the  hamiltonian 
\begin{equation}
     H={\gamma\over\pi\sin\gamma}\left(-ab-\sum_{i=1}^{N-1}e_{i}\right)
 \label{hamilt}
\end{equation}
(the normalization guarantees unit sound velocity) has been studied 
in the continuum most recently in
 \cite{Nicholsmult} where it was found  {\bf for any $a>0$}  to give rise to 
 the generating function of scaled gaps, 
 in the sector with 
 $L$ non contractible lines%
\footnote{A non contractible line is a strand propagating
throughout the system. Thus, $e_i$ acting on two non contractible lines
at $i$ and $i+1$ is zero by definition. Figure~\ref{blob} has $L=1$ non
contractible line running from the bottom to the top of the figure 
(top and bottom are later wrapped onto an annulus). Note
also that $L$ and $N$ must have the same parity.}
propagating:
\begin{equation}
K_{r,r+L}= {\rm Tr} \, q^{L_{0}-c/24}={q^{h_{r,r+L}}-q^{h_{r,-r-L}}\over q^{c/24}P(q)}
\label{Nichols}
\end{equation}
Here, as  usual, $L_0$ is a Virasoro generator,
\begin{equation}
  c = 1-\frac{6}{p(p+1)}
\label{central}
\end{equation}
 is the central charge for $\gamma = \frac{\pi}{p+1}$,  and 
\begin{equation}
  h_{r,s} = \frac{[(p+1)r-ps]^2-1}{4p(p+1)}
\label{Kac}
\end{equation}
are the conformal weights of the Kac table. Moreover, $P(q) =
\prod_{n=1}^\infty (1-q^n)$. We stress that in (\ref{Nichols}) we have
$q=\exp(2\pi i \tau)$, where $\tau$ is the standard modular parameter. As this
meaning of $q$ will be reserved for the argument of the spectrum generating
functions, no confusion should arise with the {\em other} meaning of $q$ as
the quantum group deformation parameter $q=e^{i\gamma}$ appearing in the
parameterization of $x$.

The case $r=1$ of (\ref{Nichols}) is the usual hamiltonian for the
loop model with free boundary conditions, since in this case
$y=2\cos\gamma=x$. In this case, the generating function of scaled gaps
has indeed been known for a very long time \cite{BauerSaleur} to be
given by $K_{1,1+L}$. The case $n>1$ as presented in \cite{Nicholsmult} 
appeared to be new, although it is in fact related to results in 
\cite{BauerSaleur} (and has algebraic connotations in terms of 
representation theory of the blob versus the Virasoro algebra). In this paper, we  shall discuss (\ref{Nichols})
further, interpret it in the 
language of the loop model, and correct it whenever necessary.

We stress that the independence upon $a$ is a truly remarkable phenomenon: it
can be interpreted by saying that once the algebraic structure of the
hamiltonian is decided, the continuum limit does not depend on the (boundary)
details. Another way to view this independence is the following. The $R$
matrix of the loop model with spectral parameter $u$, acting on strands $i$
and $i+1$, can be written $R_i(u) = 1 + f(u) e_i$, and $f(u) =
\frac{\sin(u)}{\sin(\gamma-u)}$ is easily found by solving the Yang-Baxter
equations. Writing similarly the boundary matrix as $B(u) = 1 + g(u) b$, the
Sklyanin (or reflection) equations relate $B(u)$ and $R_1(u)$, giving rise to
a solution for $g(u)$ (see Eq.~(40) in \cite{Doikou03}) that contains an
arbitrary constant of separation $\zeta$. The arbitrariness of $\zeta$ is
analogous to the $a$-independence discussed above.

\section{Boundary loop model}
\label{sec:BLM}

 \begin{figure}
\begin{center}
  \leavevmode
  \includegraphics[scale=.45,angle=0 ]{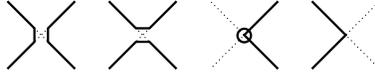}
\end{center}
\protect\caption{Vertices of the boundary loop model on the tilted square lattice. Bulk vertices are in
  any of two different states (corresponding to the generators $1$ and $e_i$).
 Left boundary vertices are always blobbed (generator $b$), and right boundary
vertices are unblobbed (generator $1$).}
\label{vertices}
\end{figure}

We now want to study the blob algebra in the context of isotropic
dense 
loop models, which are described by a transfer matrix instead of a 
hamiltonian. We recall that in the bulk, these models are defined by 
dense coverings of the (tilted) square lattice with self avoiding and mutually 
avoiding loops, each vertex allowing two possible configurations (see
Figure~\ref{vertices} and also Figures~\ref{clusters}--\ref{cexi} below),
and each loop coming with a fugacity $x$. What happens at the 
boundary is the subject of this paper. 

The sum over TL generators in the 
hamiltonian is replaced in that case by a product 
\begin{equation}
 T_{0}=\prod_{i=1}^{\lfloor (N-1)/2 \rfloor} (1+e_{2i})
 \prod_{i=1}^{\lfloor N/2 \rfloor} (1+e_{2i-1})
\label{TM}
\end{equation}
where $\lfloor \ldots \rfloor$ denotes the integer part.
By analogy we will  supplement this by boundary contributions  so the 
full transfer matrix reads
\begin{equation}
    T=(1+\lambda b)T_{0}
\label{TMlambda}
\end{equation}

In view of the $a$-independence of (\ref{Nichols}), we would expect the
critical exponents associated with $T$ to be independent of $\lambda$. This
independence is checked numerically below (see Figure~\ref{fig:conj}).
From a geometrical point of view the most natural choice is then
$\lambda=\infty$, so that after a trivial rescaling $T=bT_{0}$. This
can be interpreted as a lattice model for which {\sl every loop}
touching the boundary gets a modified weight $y$ instead of $x$. We
will now study the conformal properties of this model (which we call
the {\em boundary loop model}), as a function of $x$ and $y$.

\subsection{Conformal boundary conditions and critical exponents}

The results from \cite{BauerSaleur} and more recently 
\cite{Nicholsmult} 
suggest the following. Recall the parametrization $x = 2 \cos \gamma$; 
for each $y(r)$ we solve
(\ref{yloop}) to get $r$. This gives a {\sl real} number $r \in
(0,\frac{\pi}{\gamma})$. The leading
eigenvalue in the loop model transfer matrix with $L$ non
contractible lines should scale with conformal weight $h(y) \equiv h_{r,r+L
}$. For when $\gamma = \frac{\pi}{p+1}$ with $p$ integer and $r=1,2,\ldots,p$
integer this is a rigorous consequence of \cite{BauerSaleur} and the loop-RSOS
correspondence, as will be discussed in Section~\ref{sec:RSOS} below.

\begin{figure}
\begin{center}
\leavevmode
  \hskip-2.0cm \includegraphics[scale=.33,angle=-90 ]{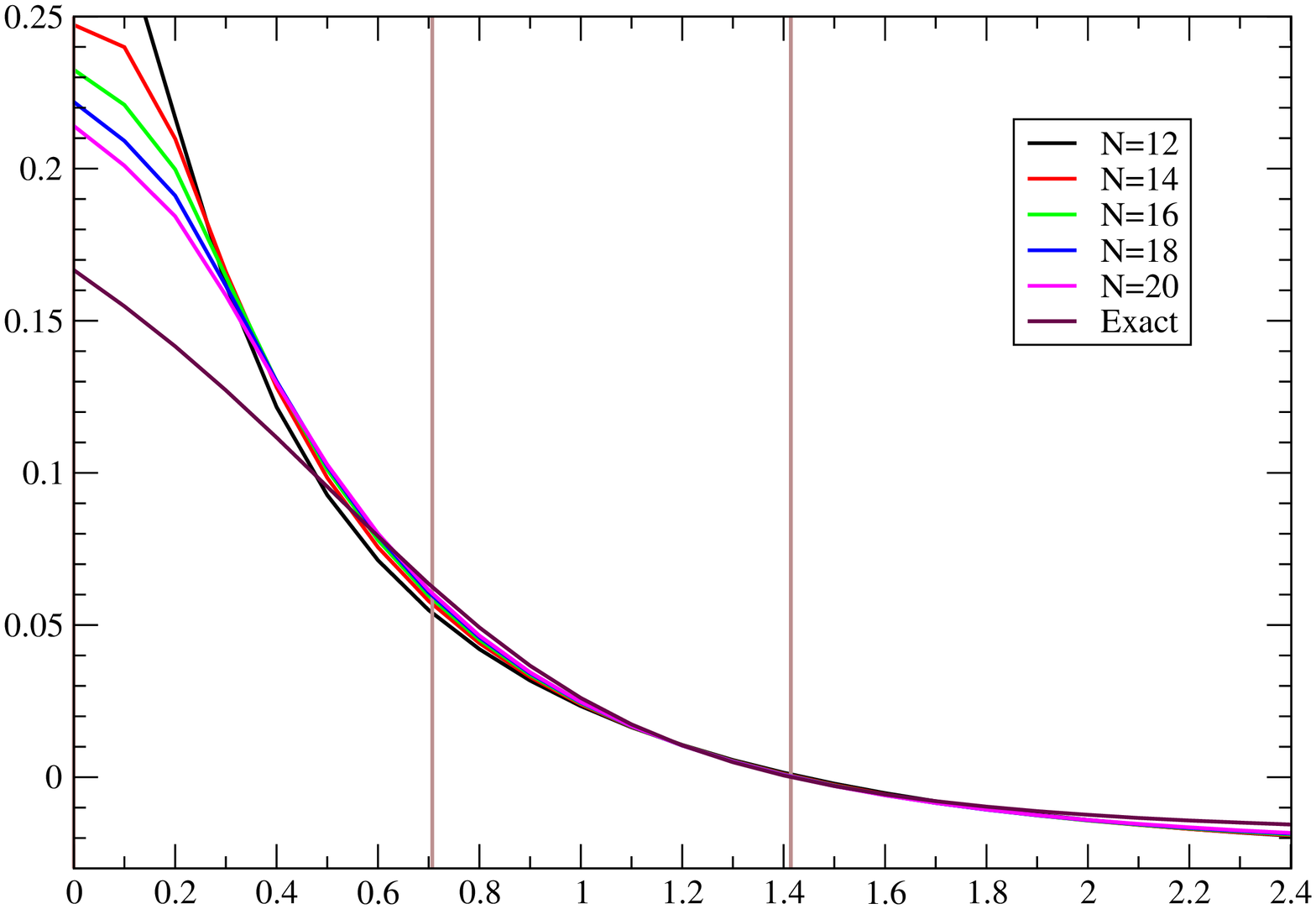} \hskip-1.0cm
  \includegraphics[scale=.33,angle=-90 ]{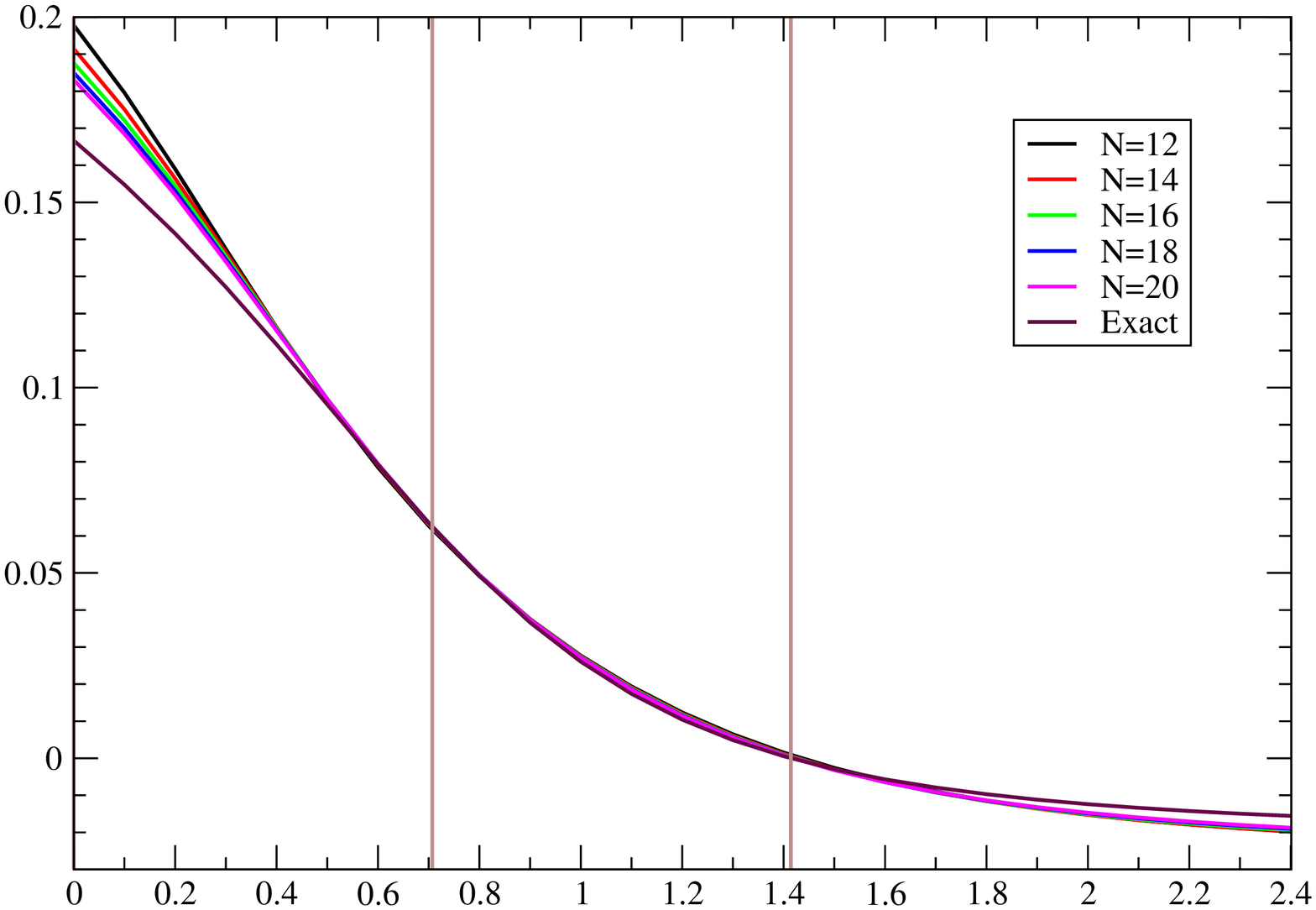} \hskip-1.0cm
\end{center}
\protect\caption{(Color online). Numerical check of the conjecture
  $h(y) = h_{r,r}$ as a function of $y$, here for $L=0$ and $p=3$ (the
  Ising model). The results were obtained from the finite-$N$
  corrections to the leading eigenvalue of the transfer matrix
  (\ref{TMlambda}), obtained by exact diagonalization techniques, for
  widths up to $N=20$. The left (resp.\ right) panel shows the choice
  $\lambda=1$ (resp.\ $\lambda=\infty$), but the extrapolated results
  ($N\to\infty$) appear to be independent of $\lambda$. The vertical
  lines represent the particular values (\ref{yloop}).}
\label{fig:conj}
\end{figure}

Figure~\ref{fig:conj} serves the double purpose of checking this
conjecture for $L=0$ and $p=3$ (the Ising model), and establishing the
independence of the exponents on the choice of $\lambda$ in
(\ref{TMlambda}). The agreement with the numerics is generally very
good, and even in regions with strong corrections to scaling (in
particular $y=0$) it should be noted that the finite-$N$ effects have
consistently a trend and an amplitude compatible with the conjectured
result in the thermodynamic limit. The choice $\lambda=\infty$ appears
to minimize the amplitude of the finite-size effects, and accordingly
we shall invariably adopt this choice for the subsequent numerical checks.

\begin{figure}
\begin{center}
\leavevmode
  \includegraphics[scale=.33,angle=-90 ]{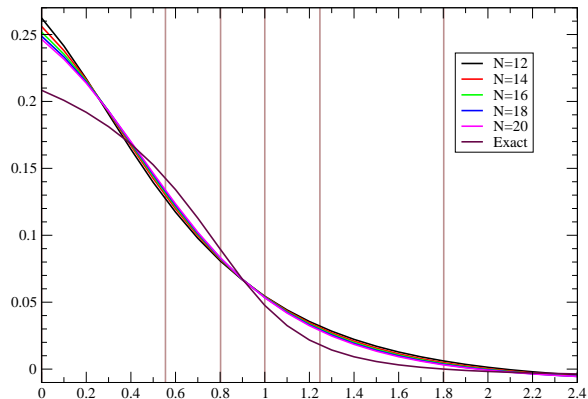}
\end{center}
\protect\caption{(Color online). Another check of the conjecture $h(y)
  = h_{r,r}$ for $L =0$, here for $p=6$ (the tricritical Potts
  model).}
\label{fig:conj2}
\end{figure}

Another check, still for $L=0$ but with a higher value $p=6$ (the
tricritical Potts model), is shown in Figure~\ref{fig:conj2}.

From the point of view of the blob algebra, the parametrization 
(\ref{yloop}) has nothing special compared with the other choice
of sign in (\ref{yloopsigned}), which we rewrite here as
\begin{equation}
    y(r')={\sin(r'-1)\gamma\over\sin r'\gamma}
    \end{equation}
Of course by using symmetries of the sine function we can write as 
well
\begin{equation}
    y(r')={\sin(p+2-r')\gamma\over \sin(p+1-r')\gamma}
\end{equation}
so we see that the associated exponent reads as well
$h_{p+1-r',p+1-r'+L}=h_{r'-1,r'-L}$, by the symmetry of (\ref{Kac}).

We now have to make things a little more precise and technical. The
sector with $L$ non contractible loops can in fact be considered from
two points of view, depending on whether or not the leftmost non
contractible loop is allowed to touch the left boundary. In Appendix A
we discuss in details the sector structure of the transfer matrix
$b T_0$ of (\ref{TM}) and show in particular that each of its eigenvalues
corresponds to a definite choice: either the leftmost contractible loop
is {\em forced} to touch the left boundary at least once, or it is {\em forbidden}
from ever doing it. We shall henceforth refer to these two cases as the
{\em blobbed} (resp.\ the {\em unblobbed}) sector.  Note that in both sectors
the contractible loops to the left of the leftmost non contractible loop may of
course still touch the left boundary.

\begin{figure}
\begin{center}
\leavevmode
  \includegraphics[scale=.33,angle=-90 ]{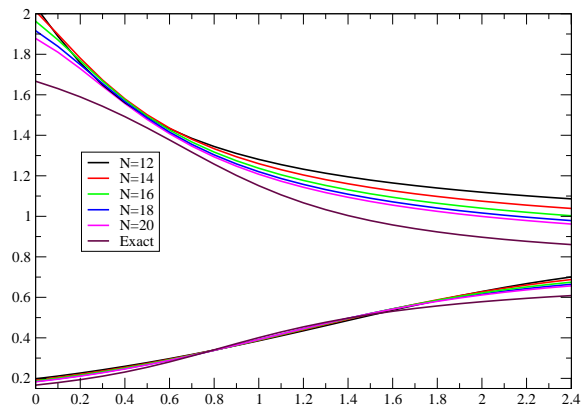}
\end{center}
\protect\caption{(Color online). Exponents $h(y)$ in the presence of $L=2$ non contractible
loops, here for $p=3$ (the Ising model). The lower (resp.\ upper) curves
show the numerical results for the (constrained) boundary loop model. They
agree with the conjecture $h(y)=h_{r,r+L}$ (resp.\ $h(y)=h_{r,r-L}$) as
claimed.}
\label{fig:L2check}
\end{figure}

We claim that these two cases both correspond to conformal boundary
conditions, which are different. For entropic reasons, the largest eigenvalue
in the blobbed sector is obviously greater than the largest eigenvalue in the
unblobbed sector. From the above arguments, the blobbed sector therefore has
the exponent $h_{r,r+L}$ indeed. The unblobbed sector meanwhile has a
different exponent $h_{r,r-L}$. This can be checked numerically, and is
illustrated in Figure~\ref{fig:L2check}. Note of course that when $L=0$ the
two results actually coincide, as they should, since the two sectors are
then identical.

\subsection{Relation to the Potts model}

The boundary loop model is closely related to the $Q=x^2$ state Potts model
at coupling (inverse temperature) $J$. Indeed, ignoring first boundary
effects, the Potts model partition function can be written as \cite{BKW76}
\begin{equation}
 Z = \sum_{\rm clusters} \left( e^J-1 \right)^B Q^C
   = Q^{V/2} \sum_{\rm loops} Q^{\ell/2}
\end{equation}
The first sum is over bond percolation clusters consisting of $B$ bonds and
$C$ connected components. The second sum is over loops on the medial lattice
that separate the clusters and their duals, with $\ell$ being the number of
loops and $V$ the number of Potts spins. We have here supposed that the model
is defined on a square lattice and stands at its critical temperature,
$e^J-1=Q^{1/2}$. The equivalence between the cluster and loop formulations is
obtained by applying the Euler relation. Note that the local configurations of
the loops correspond precisely to the first two vertices of
Fig.~\ref{vertices}.

In the sector $L=0$ we have claimed that the exponent of the boundary
loop model is $h_{r,r}$ in the parameterization (\ref{yloop}). We now
wish to check that this claim is consistent with known results on the
Potts model. To that end, consider the $Q$-state Potts model in the
same annular geometry as the boundary loop model. Denote by $P_1$ and
$P_2$ two points on the left boundary, and let boundary spins on the
interval $P_1P_2$ be constrained to take a subset of $Q_{\rm s}$
states (with $Q_{\rm s} \le Q$). In particular, when $Q_{\rm s} = 1$,
this boundary condition corresponds to the Potts spins being fixed on the
interval $P_1P_2$ and free on the remainder of the boundary. The
modified partition function reads
\begin{equation}
 Z(P_1P_2) = \sum_{\rm clusters} \left( e^J-1 \right)^B Q^C
             \left( \frac{Q_{\rm s}}{Q} \right)^{C(P_1P_2)}
           = Q^{V/2} \sum_{\rm loops} Q^{\ell/2}
             \left( \frac{Q_{\rm s}}{Q} \right)^{\ell(P_1P_2)}
\label{ZCardy}
\end{equation}
where we have used the Euler relation as before. The number of clusters
(resp.\ loops) that touch $P_1P_2$ is denoted $C(P_1P_2)$ (resp.\
$\ell(P_1P_2)$), and obviously we have $C(P_1P_2) = \ell(P_1P_2)$.
We stress that $C$ and $\ell$ still denote the {\em total} number of
clusters and loops.
Now, (\ref{ZCardy}) is a special case of the boundary loop model with the
correspondence between weights
\begin{eqnarray}
 Q &=& x^2 \nonumber \\
 Q_{\rm s} &=& x y
\end{eqnarray}

In particular, for $Q_{\rm s}=1$ we have $r=p-1$ in (\ref{yloop}). The
corresponding special case of our general claim is therefore that the
operator that changes the Potts model boundary conditions from free to
fixed is $\phi_{p-1,p-1} = \phi_{1,2}$. This indeed coincides with a
well-known result of Cardy \cite{Cardy92}.

Another verification is furnished by $Q=3$, $Q_{\rm s}=2$. The claim is
then that the operator that changes the boundary conditions from free to
mixed is $\phi_{2,2}$ with conformal weight $h_{2,2}=\frac{1}{40}$. This
is again a well-known result.

\subsection{Spectrum generating functions}

Further numerical study shows that the spectrum generating functions 
for the blobbed and unblobbed sectors of the boundary loop model
are simply the characters of generic 
irreducible representations of the Virasoro algebra, i.e., respectively,
\begin{equation}
 \begin{tabular}{rl}
   \hbox{Blobbed sector}:  & $Z_L^\ast(r)= {q^{h_{r,r+L}-c/24}\over P(q)}$ \\
  \hbox{Unblobbed sector}: &
  $Z_L(r) = {q^{h_{r,r-L}-c/24}\over P(q)}$ \\
 \end{tabular}
\label{specgenfunc}
\end{equation}
This can be related with the structure of the basis of the blob 
algebra  and the absence of truncation of the Bratelli diagram 
\cite{MartinSaleur}. The precise finite-size definition of $Z_L^\ast$ and $Z_L$  in terms of the transfer matrix
blocks $T_L^\ast$ and $T_L$ (defined in Appendix A).
is given in (\ref{app:charK})--(\ref{app:decomp1}); note that we have set $j=L/2$ in
Appendix B.

The leading behavior in these expressions defines the exponents $h_{r,r\pm L}$
and has already been checked in Figs.~\ref{fig:conj}--\ref{fig:L2check} above.
The coefficients of the terms up to level 6 in the development
\begin{equation}
 1/P(q) = 1 + q + 2q^2 + 3q^3 + 5q^4 + 7q^5 + 11q^6 + \ldots
\end{equation}
have been verified by computing the first 32 eigenvalues of $T$, for sizes
up to $N=24$, and looking for integer gaps in the spectrum of critical
exponents. For definiteness we have concentrated on the case $p=3$ and $L=2$.
Independent computations were made in the blobbed and the unblobbed sectors.
Moreover, the verification was made both for $y=x$ and for a generic value
of $y$, and in either case the absence of singular vectors up to and
including level 6 was ascertained.

It is important to stress that the spectrum generating functions are 
not given by (\ref{Nichols}) for the BLM. The full 
loop transfer 
matrix actually contains more information, but can be truncated when 
$r$ is integer.
    
\subsection{Partition function identities}
\label{sec:Z} 

We are now interested in the situation where the non contractible
loops wrap around the annulus. The question then arises of which
weight should be given to these loops (while our results for the 
exponents have some overlap with \cite{Nicholsmult}, the following has 
never appeared before). We will give in general the
weight $l$ to non contractible loops that do not touch the boundary
(i.e., the outer rim of the annulus),
and weight $m$ to those that do (there is at most one). We now claim
that the full partition functions are given by adding sectors
$Z_L^\ast(r)$ and $Z_L(r)$ with the following amplitudes. We
parametrize $l$, $m$ in terms of two numbers $\alpha$, $\beta$:
\begin{eqnarray}
	l&=&2\cosh\alpha\nonumber\\
       m&=&{\sinh (\alpha+\beta)\over\sinh \beta}
\end{eqnarray}
Then the amplitudes are
\begin{eqnarray}
    D_L^\ast &=&{\sinh(L
\alpha+\beta)\over\sinh\beta}\nonumber\\
    D_L &=&{\sinh(L
\alpha-\beta)\over\sinh(-\beta)}\label{guess}
    \end{eqnarray}

\begin{figure}
\begin{center}
\leavevmode
  \includegraphics[scale=.33]{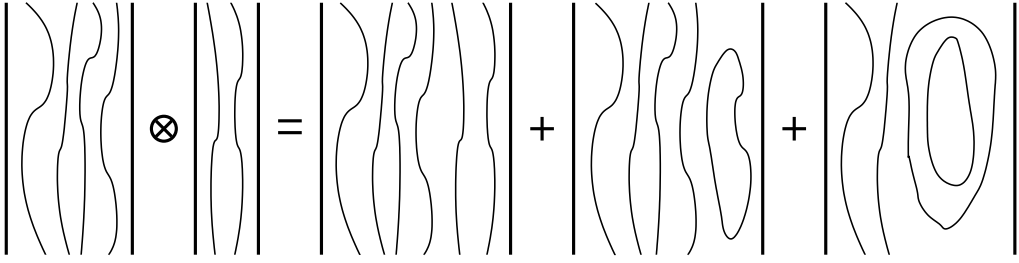}
\end{center}
\protect\caption{Pictorial representation of the recursion relation
  (\ref{recurs}).}
\label{fig:trick}
\end{figure}

A complete proof of this statement is presented in Appendix B. Alternatively,
we can argue that, because one can get from a sector $L$ to a sector $L +2$ by
adding two non contractible loops at the right of the annulus (here seen as a
periodic strip), which can or not get contracted with the first $L$ ones, the
generic amplitudes have to obey a recursion relation (which has a deep
algebraic nature, see \cite{ReadSaleur}). Considering first the amplitudes
$D_L^0$ for the simpler problem with transfer matrix $T_0$ (i.e., without the
boundary generator $b$) one has
\begin{equation}
    D_{L}^0 D_{2}^0 = D_{L}^0 + D_{L+2}^0 +D_{L-2}^0
\end{equation}
with initial values $D_0^0=1$ and $D_2^0=l^2-1$.
A pictorial rendering of this relation is shown in Figure~\ref{fig:trick}.
The solution reads explicitly $D_L^0 = U_L(l/2)$, where $U_L$ is the $L$th
Chebyshev polynomial of the second kind. Turning now to the boundary loop
model we must have similarly
\begin{eqnarray}
    D_{L} D_{2}^0 &=& D_{L} + D_{L+2} + D_{L-2} \nonumber \\
    D_{L}^\ast D_{2}^0 &=& D_{L}^\ast + D_{L+2}^\ast + D_{L-2}^\ast
\label{recurs}
\end{eqnarray}
The initial values can then be determined to be $D_{0}=D_0^\ast=1$, 
$D_{2}=l^{2}-lm-1$ and $D_{2}^\ast=lm-1$, from which the general 
formulas  (\ref{guess})
follow. In addition to the actual proof of Appendix B, we have checked
(\ref{guess}) by formal manipulations of 
transfer matrices up to size $N=6$, along the lines of \cite{SalasRing}.

\begin{figure}
  \begin{center}
    \leavevmode
    \epsfysize=60mm{\epsffile{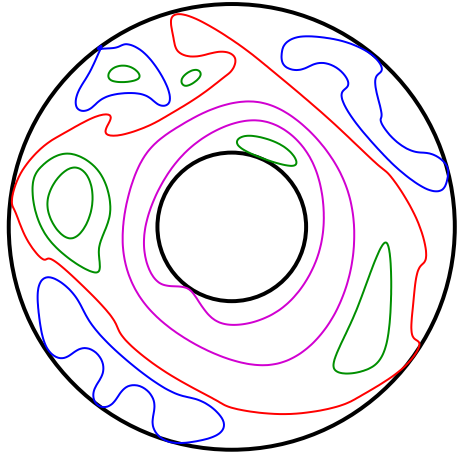}}
  \end{center}
  \protect\caption{(Color online). The most general model studied in this
    paper. We distinguish four types of self-avoiding loops on the annulus:
    contractible loops touching (blue color in the figure) or not touching
    (green) the outer rim of the annulus, and similarly for non contractible
    loops (red or purple).}
  \label{partfunct}
\end{figure}

The amplitudes can  then be used to write down the general partition 
functions in the case $N$ even (we set $L=2j$)
\begin{eqnarray}
    Z=q^{-c/24}\left[\sum_{j=0}^{\infty} 
    {\sinh(2j\alpha+\beta)\over\sinh\beta} \; {q^{h_{r,r+2j}}\over P(q)}-
    \sum_{j=1}^{\infty}{\sinh(2j\alpha-\beta)\over\sinh\beta} \;
    {q^{h_{r,r-2j}}\over P(q)}\right]\label{partfunc}
    \end{eqnarray}
This partition function correspond to the most general case 
represented in Figure~\ref{partfunct}:  contractible loops in the bulk get 
the weight $x$, those touching the boundary a weight $y$,  
 non 
contractible loops not touching the boundary  a weight $l$ and 
the others a weight $m$.  

The simplest situation occurs  when 
\begin{equation}
    m={\sinh(r+1)\alpha\over\sinh r\alpha} \,,
    \end{equation}
that is, the parameter $m$ has the same formal algebraic relationship 
with $l$ that $y$ does with $x$. Then $\beta=r\alpha$---the same $r$ as 
in (\ref{yloop})---and one can write 
\begin{eqnarray}
    Z=q^{-c/24}\left[\sum_{j=0}^{\infty} 
    {\sinh(2j+r)\alpha \over\sinh r\alpha} \; {q^{h_{r,r+2j}}\over P(q)}-
    \sum_{j=1}^{\infty}{\sinh(2j-r)\alpha\over\sinh r\alpha} \;
   { q^{h_{r,r-2j}}\over P(q)}\right]\label{partfunci}
    \end{eqnarray}
If moreover one is in a degenerate case where $r$ is integer, one can reorganize 
the sum (\ref{partfunci}) into
\begin{eqnarray}
    Z=\sum_{j=0}^{\infty} 
       {\sinh(2j+r)\alpha\over\sinh r\alpha} \; K_{r,r+2j}-q^{-c/24}
       \sum_{j=1}^{r-1}{\sinh(2j-r)\alpha\over\sinh r\alpha} \;
       {q^{h_{r,r-2j}}\over P(q)}\label{partfuncii}
       \end{eqnarray}
By pairing up terms with $j$ and $r-j$ in the second sum,
this can in turn be rewritten as
\begin{eqnarray}
    Z = \sum_{j=-[r/2]}^{\infty} 
       {\sinh(2j+r)\alpha\over\sinh r\alpha} \; K_{r,r+2j}\label{niceZ}
       \end{eqnarray}
where $\lfloor \ldots \rfloor$ denotes the integer part. When $r$ is even, the 
contribution from $j=-r/2$ actually disappears. We thus get a sum over 
irreducible representations of the Virasoro algebra.

We claim that the subtractions occurring in the partition function do 
occur in finite size as well. These subtractions involve the conformal 
weights $h_{r,r+2j}$ and $h_{r,-r-2j}$ and correspond respectively to 
the blobbed sector with $L=2j$ non contractible loops and the unblobbed sector
with $L=2j+2r$.
In other words, we claim there are level coincidences in 
finite size between the blobbed and unblobbed sectors when $r$ 
is an integer: this in fact follows from the theory of 
representations of the blob algebra. 

\begin{table}
 \begin{center}
 \begin{tabular}{l|cccccc}
 $f$             & $T_0^*$ & $T_2^*$ & $T_4^*$ & $T_6^*$ & $T_2$ & $T_4$ \\
 \hline
 -0.338946565198 & 1       &         &         &         &       &       \\
 -0.295466694605 &         & 1       &         &         &       &       \\
 -0.247750936031 & 1       &         &         &         & 1     &       \\
 -0.238365774752 &         & 1       &         &         &       &       \\
 -0.210311165453 & 1       &         &         &         &       &       \\
 -0.201367085486 &         & 1       &         &         &       &       \\
 -0.195052854163 & 1       &         &         &         & 1     &       \\
 -0.184121456296 &         & 1       &         &         &       &       \\
 -0.171718537721 & 1       &         &         &         &       &       \\
 -0.168641596700 &         &         & 1       &         &       &       \\
 -0.149149353298 &         & 1       &         &         &       &       \\
 -0.134931152674 &         &         & 1       &         &       &       \\
 -0.125859871619 & 1       &         &         &         & 1     &       \\
 -0.110023422061 & 1       &         &         &         &       &       \\
 -0.097996769822 &         & 1       &         &         &       &       \\
 -0.085859268861 &         & 1       & 1       &         &       & 1     \\
 -0.069318933352 & 1       &         &         &         & 1     &       \\
 -0.067354247176 &         & 1       &         &         &       &       \\
 -0.048891555720 & 1       &         &         &         & 1     &       \\
 -0.041781216253 &         & 1       &         &         &       &       \\
 -0.036787385047 &         &         & 1       &         &       &       \\
 -0.027592998174 & 1       &         &         &         &       &       \\
 -0.012286435222 &         & 1       &         &         &       &       \\
 -0.003076941021 &         &         & 1       &         &       &       \\
 -0.000000000000 &         &         &         & 1       &       &       \\
 \end{tabular}
 \end{center}
 \caption{Complete set of levels for the various transfer matrix blocks
   (cf.~Appendix A) in the case of $N=6$ strands, for $p=6$ (the tricritical
   Potts model) and $r=1$ (i.e., $y=x$). The left column shows 
   $-N^{-1} \log \lambda$ (rounded to 12 digits), where $\lambda \ge 1$
   is the transfer matrix eigenvalue, and the remaining columns show the
   multiplicities of $\lambda$ within the various blocks. A blank entry
   denotes zero multiplicity. The level coincidences mentioned in the
   text are clearly observed. (There is one extra coincidence of the level
   $-0.085859268861$, between $T_2^*$ and $T_4^*$.)}
 \label{tab:levels}
\end{table}

To be more precise, one can make sense of expressions such as (\ref{niceZ})
in finite size by replacing the definition (\ref{Nichols}) of  the characters
$K_{r,s}$ by traces of transfer matrix blocks, as outlined in Appendix B.
Care should then be taken that the
annulus is wide enough to accommodate the prescribed number of non contractible
lines, which amounts to replacing the upper limit in the summation
(\ref{niceZ}) by $N/2$.
A numerical check of the level coincidences is shown in
Table~\ref{tab:levels}, for the case $N=6$, $p=6$ and $r=1$ (i.e.,
$y=x$). We find indeed that the level spectrum of the transfer 
matrix (see Appendix A)
block $T_{2j+2r}$ is a {\em proper subset} of that of $T_{2j}^\ast$, for
all allowed values of $j$ (i.e., $j=0,1,\ldots,N/2-2r-1$). We have checked
this statement for several other values of $N$, $p$ and $r$, including on
examples involving many more levels.

Although we have restricted to $N$ even, similar results can be 
written for $N$ odd, with the difference that $L$ is odd (and thus 
never vanishes). The generating function then reads
\begin{eqnarray}
    Z=q^{-c/24}\left[\sum_{j=0}^{\infty} 
    {\sinh[(2j+1)\alpha+\beta]\over\sinh\beta}{q^{h_{r,r+2j+1}}\over P(q)}-
    \sum_{j=0}^{\infty}{\sinh[(2j+1)\alpha-\beta]\over\sinh\beta}
    {q^{h_{r,r-2j-1}}\over P(q)}\right]\label{partfuncodd}
    \end{eqnarray}

\subsection{Neumann  boundary conditions}
\label{sec:Neumann}

\begin{figure}
\begin{center}
	\leavevmode
	\epsfysize=60mm{\epsffile{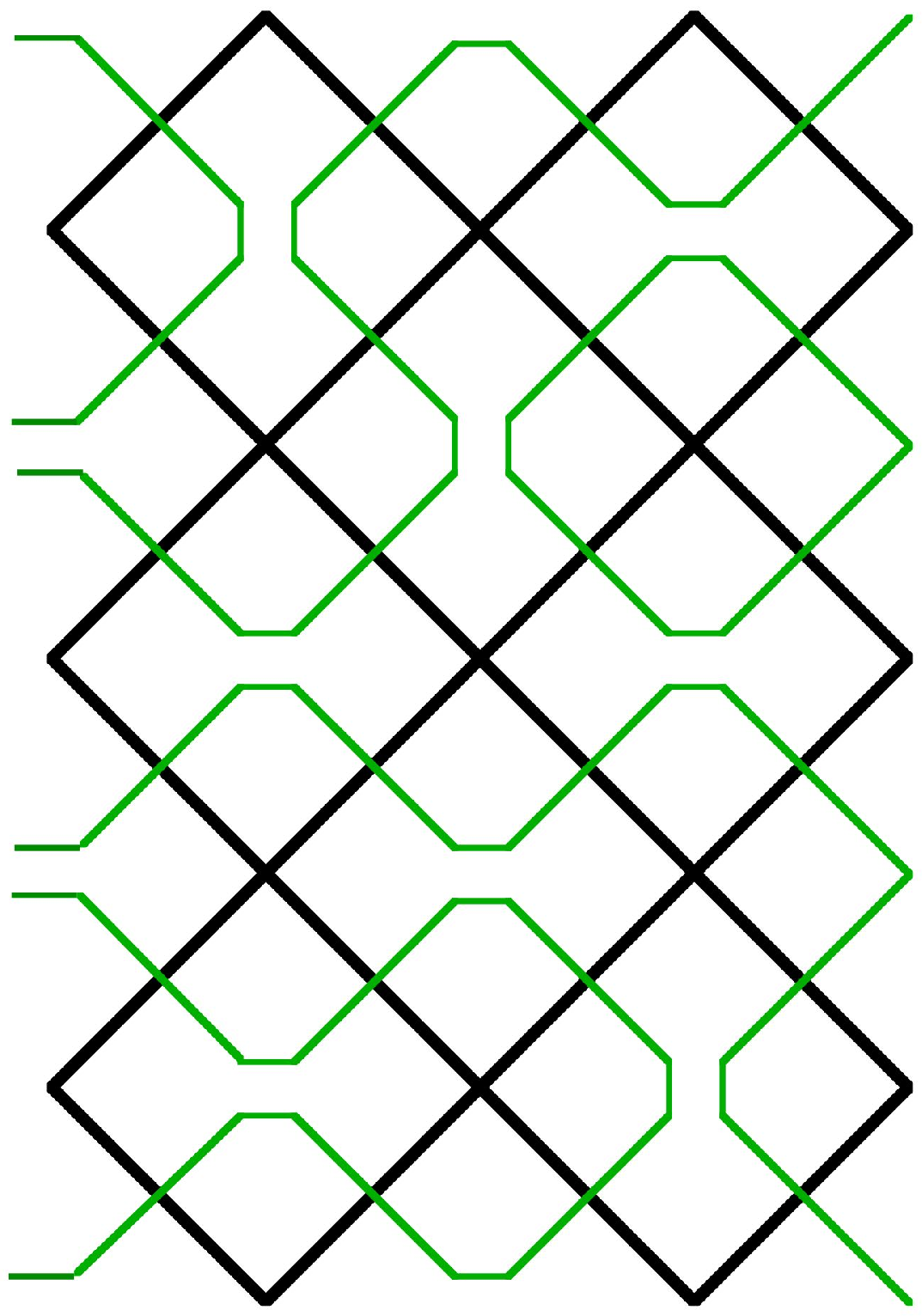}}
\end{center}
\protect\caption{(Color online). Configuration in which the two boundaries of
  the annulus (here shown as a periodic strip) have respectively Dirichlet and
  Neumann boundary conditions. RSOS faces are shown in black, and the loops
  that separate clusters at constant height are green. The open lines going
  from the left to the left boundary carry unit weight.}
\label{opencluster}
\end{figure}

\begin{figure}
\begin{center}
\leavevmode
\epsfysize=60mm{\epsffile{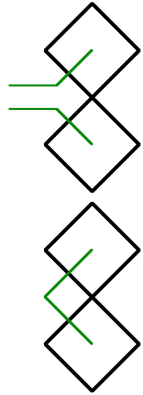}}
\end{center}
\protect\caption{(Color online.) Two possible interactions between the loops and the left
boundary: fork (top image) and blob (bottom image).}
\label{Don}
\end{figure}

If we go back to the derivation of the partition function for the 
$O(n)$ model \cite{Nienhuis} we see that Dirichlet boundary conditions for the $O(n)$ variable will 
translate into having a loop extremity on every point of the boundary, and 
the open lines thus obtained get a weight one (since the $O(n)$ 
variable is fixed to say $\vec{S}=(1,0,\ldots 0)$ on the boundary.)  
Let us assume this carries over to the fully packed case and our 
geometry (see Figure~\ref{opencluster}) , where we thus demand that every point on  the boundary looks 
like the top diagram on Figure~\ref{Don}, which we call a fork. 
We also require that 
open loops thus formed all carry a weight unity. It is then easy to see 
that this is equivalent to marking loops touching the boundary with a 
blob having parameter $y=1$. In turn, with  the usual parametrization 
for $x$  we get the associated conformal weight to be
\begin{equation}
    h_{\rm Twist}=h_{p/2,p/2}={p^{2}-4\over 16 p(p+1)}\label{twist}
    \end{equation}
since $y=1$ in (\ref{yloop}) when $r=p/2$.
This corresponds to a twist operator, or the dimension of the boundary 
field changing boundary conditions from Neumann to Dirichlet.

We note that   this dimension of the twist operator agrees with the 
one proposed in \cite{Kostov} after some reinterpretation of the 
results. The formula given in this reference is slightly different
\begin{equation}
    h_{\rm Twist}^{\rm K}=h_{{p+1 \over 2},{p+1 \over 2}}={(p+1)^{2}-4\over 16 p(p+1)}
    \end{equation}
but turns out to hold for the dilute phase of the $O(n)$ model \cite{Nienhuis}
only \cite{Kostovprivcomm}. For the dense case, 
it has to be replaced by (\ref{twist}).

We can now write the partition function with Dirichlet boundary 
conditions on one rim of the annulus and Neumann on the other, simply by setting 
$r={p\over 2}$ in our formulas. An interesting limit to consider is 
then $p\rightarrow\infty$, $m=1,l=2$ 
where we find from (\ref{partfunc})
\begin{equation}
    Z=q^{-1/24}\sum_{j=-\infty}^{\infty}
     {q^{(2j+1)^{2}/16}\over P(q)}
    \end{equation}
which is  the usual Dirichlet-Neumann partition 
function for the free boson. The presence of the two kinds of terms 
$h_{r,r\pm L}$ in (\ref{partfunc}) is crucial to recover this limit. 

\begin{figure}
\begin{center}
\leavevmode
  \includegraphics[scale=.33,angle=-90 ]{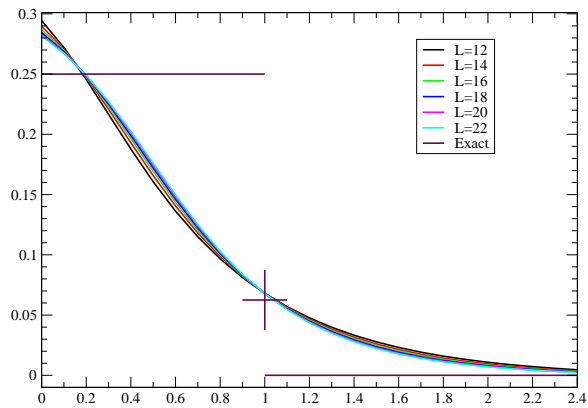}
\end{center}
\protect\caption{(Color online). Exponents $h(y)$ with $L=0$ non contractible
loops, in the limit $p \to \infty$. The exact result is a step function, passing through the value $h(1) = \frac{1}{16}$ (shown as a big cross).}
\label{fig:Binf}
\end{figure}

Note that in the limit $p\rightarrow\infty$, the exponent $h(y)$ becomes a 
step function:
\begin{equation}
 h(y) = \left \lbrace \begin{array}{ll}
 \frac{(L
-1)^2}{4}    & \mbox{for } y < 1 \\
 \frac{(2L
-1)^2}{16} & \mbox{for } y = 1 \\
 \frac{L
^2}{4}       & \mbox{for } y > 1
 \end{array} \right.
\end{equation}
The numerical check, shown in Figure~\ref{fig:Binf}, is compatible with
this behavior, although the finite-size corrections are of course
large near the step.

 Note that in general blob and fork do not coincide: this is true 
 only when the associated weights are equal to unity. One can play 
 the game of defining a fork algebra such that the top diagram in 
 Figure~\ref{Don} defines the fork operator $f$. Required relations to 
 give to every open loop a weight $z$ are then obviously
 \begin{eqnarray}
     f^{2}=zf\nonumber\\
     e_{1}fe_{1}=ze_{1}
     \end{eqnarray}
But, by a rescaling $f\rightarrow f/z$ we get the same relations as 
the blob  relations with $y=1$!

\section{Relation to RSOS models}
\label{sec:RSOS}

\begin{figure}
\begin{center}
     \leavevmode
     \epsfysize=60mm{\epsffile{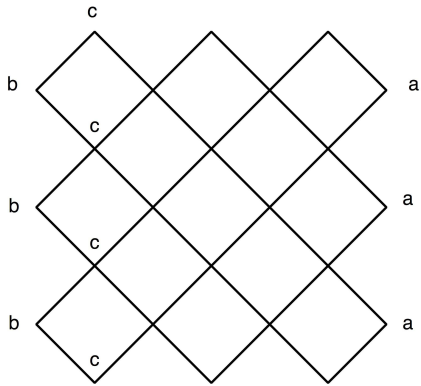}}
     \end{center}
     \protect\caption{RSOS model in an annular geometry, for $N=6$. Heights on
     the rightmost and the {\sl two} leftmost layers are fixed as
     shown. Time flows vertically, and there are periodic boundary
     conditions in the time direction.}
\label{RSOSconf}
\end{figure}

We now want to tackle the boundary loop model by another route. Recall
that in the numerical studies of Saleur and Bauer \cite{BauerSaleur}
it was found that for $A_{p}$ RSOS models [with central charge (\ref{central})]
the annulus partition function is exactly the
character $\chi_{da}$ when the following boundary conditions are
imposed: all heights on the right boundary of the annulus are fixed to
$a$ (Dirichlet boundary conditions) while on the left boundary, the
heights on the boundary are fixed to $b$ and those in the layer {\sl
next to the boundary} are fixed to $c$, with $d={\rm min}(b,c)$.%
\footnote{In order to respect the use of $a$, $b$ and $c$ in
  Ref.~\cite{BauerSaleur}, note that throughout Section~\ref{sec:RSOS},
 $a$ is not the parameter of (\ref{hamilt}), $b$ is not the blob generator
 in (\ref{blobalg}), and $c$ is not the central charge (\ref{central}).}
Note that without the
constraint on the next-to-leftmost heights, {\sl each height} in that
layer may take either of the values $b+1$ and $b-1$.  This is
illustrated in Figure \ref{RSOSconf}.

Let us now see how this choice of boundary conditions translates in
the loop model. To do so, we first note that if the loop model is
defined on $N$ strands, a time slice of the RSOS model is defined by
$N+1$ heights.  A state of the model is a collection of those heights,
denoted $|l_1,l_2,\ldots,l_{N+1} \rangle$, and we have the RSOS
constraint $l_i=1,2,\ldots,p$ with $|l_{i+1}-l_i| = 1$.
Figure~\ref{RSOSconf} corresponds to a case of even $N$. We will 
restrict to this case to start. We will also restrict to the case 
where $c=b+1$ so $b=d$, $c=d+1$; note that then $d-a$ and $N$ have the same parity.

\begin{figure}
    \begin{center}
	 \leavevmode
	 \epsfysize=30mm{\epsffile{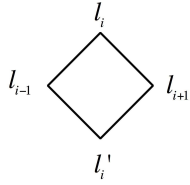}}
	 \end{center}
	 \protect\caption{Labelling of RSOS heights around a lattice face.}
	\label{Face}
	\end{figure}

We next recall briefly how the Temperley 
Lieb generators act in the RSOS representation.
Taking the time to flow downwards in Figure~\ref{Face}, the
generator $e_{i}$ acts as
\begin{equation}
    e_{i} \, \left| l_{1},\ldots,l_{i-1},l_{i},l_{i+1},\ldots,l_{N+1}
    \right\rangle =   \delta(l_{i-1},l_{i+1})\sum_{l'_{i}}
    {(S_{l_{i}}S_{l'_{i}})^{1/2}\over 
    S_{l_{i-1}}} \, \left| l_{1},\ldots,l_{i-1},l'_{i},l_{i+1},\ldots,l_{N+1}
    \right\rangle
\label{eiRSOS}
\end{equation}
where the $S_l$ are the components of the Perron-Frobenius eigenvector of
the adjacency matrix of $A_p$, and read explicitly
\begin{equation}
    S_{l}=[l]_{q} = {\sin \left( \frac{l\pi}{p+1} \right) \over
    \sin \left( \frac{\pi}{p+1} \right)}
 \label{Sl}
\end{equation}
where we have introduced the $q$-deformed numbers $[l]_q \equiv
(q^l-q^{-l})/(q-q^{-1})$.
 
The transfer matrix of the RSOS model has the usual form (\ref{TM}),
but now in terms of the $e_i$ defined by (\ref{eiRSOS}). The graphical
expansion of the partition function is obtained by taking, for each
elementary face transfer matrix $1+e_i$, either the identity---in
which case a vertical bar is drawn diagonally across the face, indicating that $l_{i}=l'_{i}$ for
this term---or the Temperley Lieb generator $e_j$, in which case a
horizontal bar is drawn, indicating that $l_{i-1}=l_{i+1}$. One gets in
this way clusters of constant heights, which can be separated by non
intersecting loops drawn on the dual lattice.

The analysis of the weights needs some modification as compared to
Pasquier's original treatment \cite{Pasquier} of the RSOS model on a
torus, but some of the key elements can be taken over. We first
consider the case of $l_1 = l_{2N+1} \equiv d$; the heights $l_2$ are
also fixed and we shall assume $l_2=d+1$ (the case $l_2=d-1$ being
similar). We further assume that there is at least one cluster
connecting the left and right boundary of the annulus (in the limit
where the aspect ratio $\rho = M/N \to \infty$ this is almost surely
the case). There is then $L
=0$ non contractible lines. Since all
clusters connecting the two boundaries have the same height $d$, we
might just as well identify them. In the graphical expansion this can
be represented by adding extra vertical bars on the two boundaries.

With this identification it follows that any loop is at the junction
between exactly {\sl two} distinct clusters. Now orient every loop in
the clockwise direction. When traversing any loop along this
direction, the cluster to its right is said to be surrounded by the
loop, whereas the cluster to its left is said to surround the loop.
Let us now represent each distinct cluster by a node, and each loop by
a directed link going from the cluster that it surrounds towards the
cluster that it is surrounded by. This then defines a directed rooted
tree, where every link is oriented towards the root, the root being the {\sl
unique} cluster connecting the two boundaries.

The weight appearing in (\ref{eiRSOS}) is then distributed on
individual loop turns as follows: Consider a loop that surrounds a
cluster at height $l$ and that is surrounded by a cluster at height
$k$. When making a right (resp.\ left) turn whilst being tangential to
a {\sl horizontal} bar, the loop picks up a factor $(S_l/S_k)^{1/2}$
(resp.\ $(S_l/S_k)^{-1/2}$). Turns being tangential to a vertical bar
do not carry any weight. The complete loop, being clockwise, makes
four more right than left turns, but only two of those excess turns
carry any weight, so the complete weight is $S_l/S_k$.

The directed tree is now undone by summing over the heights of its
nodes.  We start at a leave node. For any fixed height $k$ of the node
adjacent to the leaf, the height of the leaf node can be
$l=k\pm 1$. Summing over $l$ gives a weight%
\footnote{This is true even when $l=1$ or $l=p$, since $S_0=S_{p+1}=0$
from (\ref{Sl}).}
$(S_{k+1}+S_{k-1})/S_k = 2
\cos \left( \frac{\pi}{p+1} \right)$---note that this is {\sl
independent} of $k$---which can be attributed to the loop represented
by the link directed away from the leaf. We then remove the leaf and
its outgoing link, and proceed iteratively, moving always from
the leaves and towards the root. A special case occurs when summing
over the heights of those nodes adjacent to the root whose outgoing
link corresponds to a loop that touches the left boundary. Indeed, the
heights of those nodes have been fixed to $d+1$, and we obtain then
the weight $S_{d+1}/S_d = [d+1]_q/[d]_q$. Finally, when the whole tree has
been undone, only the root node remains, but since its height is fixed
it contributes no additional weight.

Consider now the loop model with weights $x=2 \cos \left(
\frac{\pi}{p+1} \right)$ and $y = [d+1]_q/[d]_q$---i.e., setting $r=d$ in
(\ref{yloop})---and no non contractible lines, $L
 = 0$. Its configurations
are in one-to-one correspondence with those of the RSOS model treated above
and the weights $x,y$ are the same. The spectrum of $T_{\rm loop}$ therefore
contains the complete spectrum of $T_{\rm RSOS}$, and we have verified this
numerically. In particular the leading eigenvalues of these transfer matrices
coincide. However, $T_{\rm loop}$ also contains eigenvalues not present in
$T_{\rm RSOS}$. This was to be expected, since the loop model contains
non-local information not present in the RSOS model (allowing in particular
its definition for non-integer $p$).

\begin{figure}
    \begin{center}
	 \leavevmode
	 \epsfysize=60mm{\epsffile{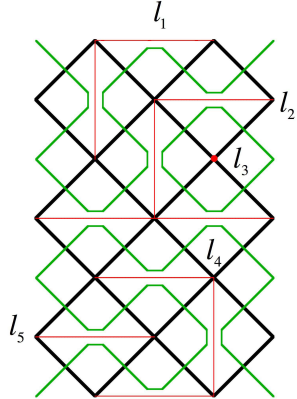}}
	 \end{center}
	 \protect\caption{(Color online.) Configuration of the RSOS clusters (in red
     color) for a case where $a=d$. Periodic boundary conditions are imposed
     in the time (vertical) direction. }
	\label{clusters}
	\end{figure}

An example of the loop-RSOS equivalence is illustrated in
Figure~\ref{clusters}, still for the case $d=a$. Denoting by
$l_{1},\ldots,l_{5}$ the heights of the five clusters%
\footnote{Note that in this example the subscripts are used differently
than in (\ref{eiRSOS}).}
(labelled from top to bottom) we find the overall weight (for
simplicity we denote $[l]_q \equiv l$): %
 \begin{equation}
     {(l_{2}l_{5})^{1/2}\over l_{1}}{(l_{1}l_{3})^{1/2}\over 
     l_{2}}{(l_{1}l_{4})^{1/2}\over l_{2}}{(l_{3}l_{4})^{1/2}\over l_{2}}
     {(l_{2}l_{5})^{1/2}\over l_{4}}{l_{4}\over l_{5}}={l_{3}\over 
     l_{2}}{l_{4}\over l_{2}}
     \end{equation}
Note that $l_4=l_1$ by the periodic boundary conditions in the time
direction.  Each of the two factors on the right-hand side is
associated with a loop.  Since $l_{3}$ can take values $d\pm 1$ while
$l_{2}=l_5=d$ and $l_{4}=l_1=d+1$, the overall weight is
$2\cos\left({\pi\over p+1}\right) \times {S_{d+1}\over S_{d}} = x y$
as claimed.
 
\begin{figure}
\begin{center}
 \leavevmode
 \epsfysize=60mm{\epsffile{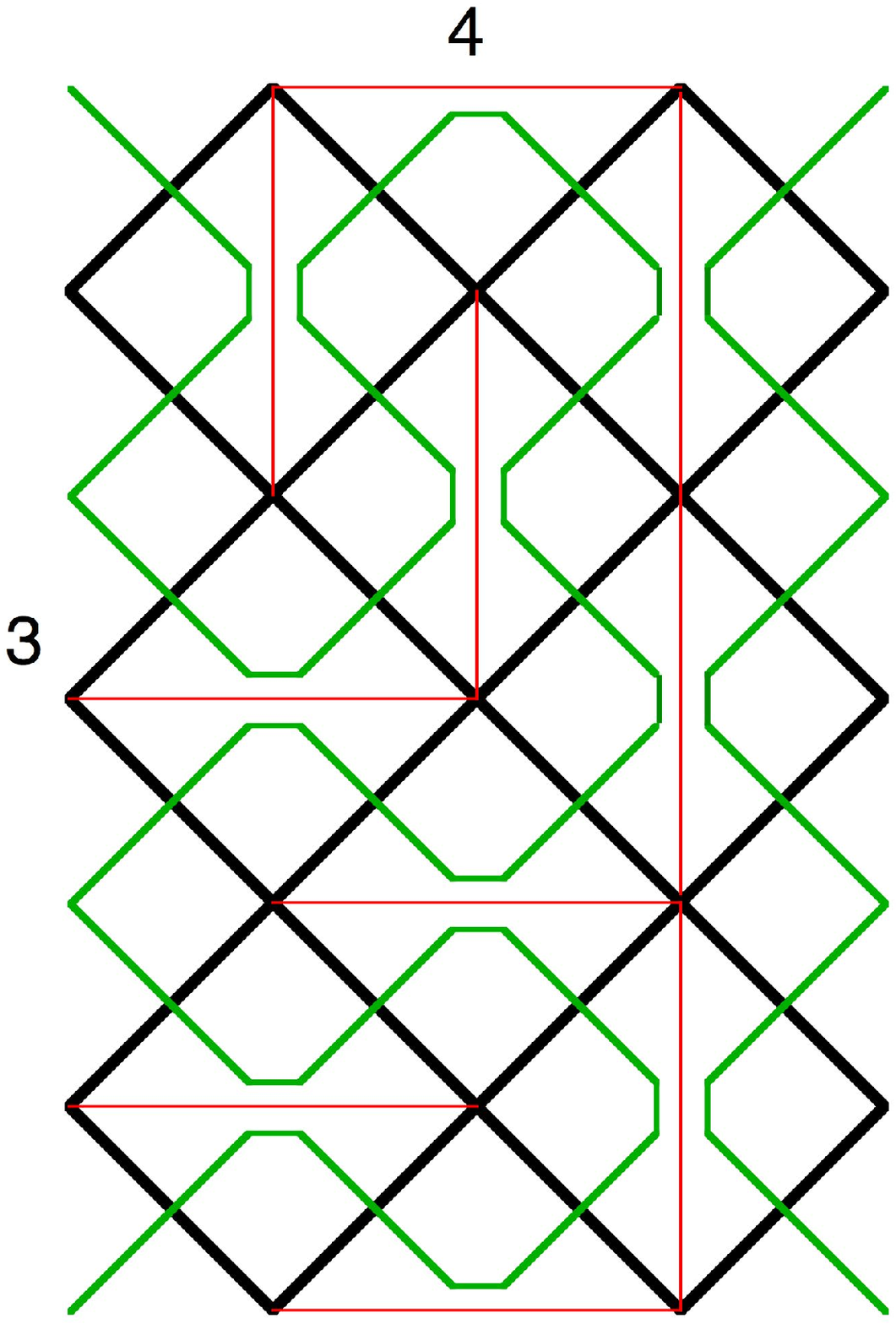}}
 \epsfysize=60mm{\epsffile{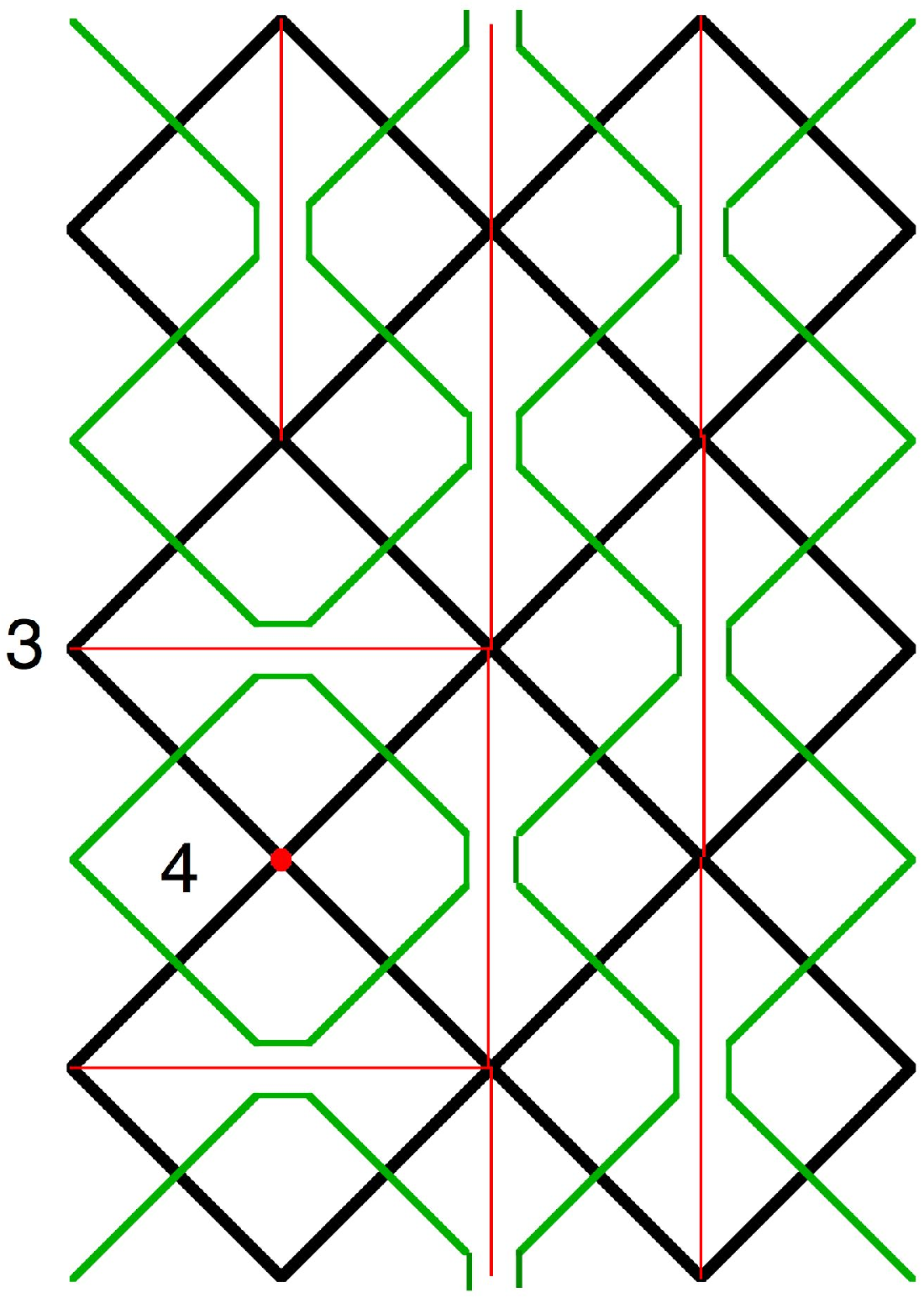}}
\end{center}
\protect\caption{(Color online.) The choice $a=3,b=4,c=1$ is not compatible with the
  configuration of loops shown on the left panel (even though there
  are two non contractible loops). It is however compatible with the
  configuration shown on the right panel.}
\label{cexi}
\end{figure}

While this construction clearly works when there are no non contractible
lines running through the annulus (ie, $a=d$), things are more
delicate when there are. Figure~\ref{cexi} shows for
instance that a configuration with two non contractible lines does not
necessarily contribute to $d=3$, $a=1$. A little thought suggests to 
simply eliminate non contractible loops touching the boundary in this 
case. 

We have indeed checked numerically that when $d-a$ is negative, the 
eigenvalues of the RSOS model are all found in the loop transfer 
matrix for the blobbed sector, and  
the leading eigenvalues coincide. Meanwhile if $d-a$ is positive, this is true provided one 
considers instead the 
unblobbed sector for the loop model. In both cases one needs
to have  $L=d-a$ and $y=y(d)$.

This corresponds simply to accepting configurations such as the one on
the right of Figure~(\ref{cexi}), but not the one on the left.

Of course the RSOS model contains more situations: one can decide to 
have $c=b-1$ instead, or to have $N$ odd. The complete set of rules 
is as follows:
\begin{equation}
 \begin{tabular}{llll}
   \mbox{$d-a > 0$ and has same parity as 
    $N$}: & $L=|d-a|$,   & $y=y(d)$,   & \mbox{unblobbed} \\
   \mbox{$d-a \le 0$ and has same parity as  
    $N$}: & $L=|d-a|$,   & $y=y(d)$,   & \mbox{blobbed} \\
   \mbox{$d-a \ge -1$ and has opposite parity of 
    $N$}: & $L=|d+1-a|$, & $y=y(p-d)$, & \mbox{blobbed} \\
   \mbox{$d-a < -1$ and has opposite parity of 
    $N$}: & $L=|d+1-a|$, & $y=y(p-d)$, & \mbox{unblobbed} \\
 \end{tabular}
\label{RSOSrules}
\end{equation}

\begin{table}
 \begin{center}
 \begin{tabular}{lr|ccccc|cccc|cc|ccc|cc}
 $f$ & $(d,a)$ & $T_0^\ast$ & & & & &
                                $T_2^\ast$ & & & & 
                                $T_4^\ast$ & & 
                                $T_2$ &  &  & $T_4$ &
 \\
& $r$ & 1 & 2 & 3 & 4 & 5 & 1 & 2 & 3 & 4 & 1 & 2 & 3 & 4 & 5 & 1 & 5
\\ \hline 
-0.338946565198 & (1,1) &1& & & & & & & & & & & & & & & \\
-0.324025479536 & (2,2) & &1& & & & & & & & & & & & & & \\
-0.316673567023 & (3,3) & & &1& & & & & & & & & & & & & \\
-0.310464485251 & (2,3) & & & &1& & & & & & & & & & & & \\
-0.302390277235 & (1,2) & & & & &1& & & & & & & & & & & \\
-0.295466694605 & (1,3) & & & & & &1& & & & & & & & & & \\
-0.291196112101 & (2,4) & & & & & & &1& & & & & & & & & \\
-0.289510241627 & (3,2) & & & & & & & &1& & & & & & & & \\
-0.288254559167 & (2,1) & & & & & & & & &1& & & & & & & \\
-0.241054964628 & (2,2) & &1& & & & & & & & & & & & & & \\
-0.238365774752 & (1,3) & & & & & &1& & & & & & & & & & \\
-0.237748281915 & (3,3) & & &1& & & & & & & & & & & & & \\
-0.234731074254 & (2,3) & & & &1& & & & & & & & & & & & \\
-0.230275094022 & (2,4) & & & & & & &1& & & & & & & & & \\
-0.230197803440 & (1,2) & & & & &1& & & & & & & & & & & \\
-0.226587692821 & (3,2) & & & & & & & &1& & & & & & & & \\
-0.223700268691 & (2,1) & & & & & & & & &1& & & & & & & \\
-0.220405235066 & (3,1) & & & & & & & & & & & &1& & & & \\
-0.212046452393 & (2,5) & & & & & & & & & & & & &1& & & \\
-0.210311165453 & (1,1) &1& & & & & & & & & & & & & & & \\
-0.201367085486 & (1,3) & & & & & &1& & & & & & & & & & \\
-0.200352797500 & (1,4) & & & & & & & & & & & & & &1& & \\
-0.194920765586 & (2,2) & &1& & & & & & & & & & & & & & \\
-0.191048760878 & (2,4) & & & & & & &1& & & & & & & & & \\
-0.187939781596 & (2,2) & &1& & & & & & & & & & & & & & \\
-0.186511467791 & (3,2) & & & & & & & &1& & & & & & & & \\
-0.186351229891 & (3,3) & & &1& & & & & & & & & & & & & \\
-0.184916564626 & (3,3) & & &1& & & & & & & & & & & & & \\
-0.184121456296 & (1,3) & & & & & &1& & & & & & & & & & \\
-0.183007717844 & (2,1) & & & & & & & & &1& & & & & & & \\
-0.182528404723 & (2,3) & & & &1& & & & & & & & & & & & \\
-0.179523463630 & (1,2) & & & & &1& & & & & & & & & & & \\
-0.178529556360 & (2,3) & & & &1& & & & & & & & & & & & \\
-0.176928229253 & (2,4) & & & & & & &1& & & & & & & & & \\
-0.172266060806 & (3,2) & & & & & & & &1& & & & & & & & \\
-0.171718537721 & (1,1) &1& & & & & & & & & & & & & & & \\
-0.170284900502 & (3,1) & & & & & & & & & & & &1& & & & \\
-0.168641596700 & (1,5) & & & & & & & & & &1& & & & & & \\
-0.167982529467 & (2,6) & & & & & & & & & & &1& & & & & \\
-0.163038059540 & (2,5) & & & & & & & & & & & & &1& & & \\
-0.155628452273 & (2,2) & &1& & & & & & & & & & & & & & \\
-0.153242699723 & (1,4) & & & & & & & & & & & & & &1& & \\
-0.149149353298 & (1,3) & & & & & &1& & & & & & & & & & \\
-0.147179401602 & (3,3) & & &1& & & & & & & & & & & & & \\
-0.140243221820 & (2,4) & & & & & & &1& & & & & & & & & \\
-0.139719837688 & (2,3) & & & &1& & & & & & & & & & & & \\
-0.135157811358 & (3,2) & & & & & & & &1& & & & & & & & \\
-0.134931152674 & (1,5) & & & & & & & & & &1& & & & & & \\
-0.132209035214 & (2,6) & & & & & & & & & & &1& & & & & \\
-0.115345804299 & (2,2) & &1& & & & & & & & & & & & & & \\
 \end{tabular}
 \end{center}
\end{table}

\begin{table}
 \begin{center}
 \begin{tabular}{lr|ccccc|cccc|cc|ccc|cc}
 $f$ & $(d,a)$ & $T_0^\ast$ & & & & &
                                $T_2^\ast$ & & & & 
                                $T_4^\ast$ & & 
                                $T_2$ &  &  & $T_4$ &
 \\
& $r$ & 1 & 2 & 3 & 4 & 5 & 1 & 2 & 3 & 4 & 1 & 2 & 3 & 4 & 5 & 1 & 5
\\ \hline 
-0.110023422061 & (1,1) &1& & & & & & & & & & & & & & & \\
-0.109559975348 & (3,3) & & &1& & & & & & & & & & & & & \\
-0.105095398499 & (3,1) & & & & & & & & & & & &1& & & & \\
-0.104419222242 & (2,3) & & & &1& & & & & & & & & & & & \\
-0.099484992148 & (2,5) & & & & & & & & & & & & &1& & & \\
-0.097996769822 & (1,3) & & & & & &1& & & & & & & & & & \\
-0.097573977039 & (1,2) & & & & &1& & & & & & & & & & & \\
-0.095502814093 & (2,2) & &1& & & & & & & & & & & & & & \\
-0.092050608093 & (1,4) & & & & & & & & & & & & & &1& & \\
-0.088269162793 & (3,3) & & &1& & & & & & & & & & & & & \\
-0.086753609983 & (2,4) & & & & & & &1& & & & & & & & & \\
-0.085859268861 & (1,5) & & & & & &1& & & &1& & & & &1& \\
-0.085433490626 & (2,4) & & & & & & &1& & & & & & & & & \\
-0.085246090691 & (3,2) & & & & & & & &1& & & & & & & & \\
-0.085077485734 & (2,1) & & & & & & & & &1& & & & & & & \\
-0.082112734265 & (2,3) & & & &1& & & & & & & & & & & & \\
-0.080981552646 & (3,2) & & & & & & & &1& & & & & & & & \\
-0.080291708506 & (2,6) & & & & & & & & & & &1& & & & & \\
-0.067354247176 & (1,3) & & & & & &1& & & & & & & & & & \\
-0.066058643675 & (3,3) & & &1& & & & & & & & & & & & & \\
-0.064397781477 & (2,3) & & & &1& & & & & & & & & & & & \\
-0.061125554776 & (1,2) & & & & &1& & & & & & & & & & & \\
-0.059945596252 & (2,4) & & & & & & &1& & & & & & & & & \\
-0.056741093622 & (3,2) & & & & & & & &1& & & & & & & & \\
-0.054049770890 & (2,1) & & & & & & & & &1& & & & & & & \\
-0.052888923290 & (2,5) & & & & & & & & & & & & &1& & & \\
-0.044940510316 & (1,4) & & & & & & & & & & & & & &1& & \\
-0.041781216253 & (1,3) & & & & & &1& & & & & & & & & & \\
-0.037564796511 & (2,2) & &1& & & & & & & & & & & & & & \\
-0.036787385047 & (1,5) & & & & & & & & & &1& & & & & & \\
-0.036787385047 & (1,6) & & & & & & & & & & & & & & & & 1 \\
-0.033076470754 & (2,4) & & & & & & &1& & & & & & & & & \\
-0.032439276523 & (3,3) & & &1& & & & & & & & & & & & & \\
-0.030419378593 & (2,6) & & & & & & & & & & &1& & & & & \\
-0.028358275153 & (3,2) & & & & & & & &1& & & & & & & & \\
-0.027839432948 & (2,3) & & & &1& & & & & & & & & & & &  \\
-0.027592998174 & (1,1) &1& & & & & & & & & & & & & & & \\
-0.020938336515 & (3,1) & & & & & & & & & & & &1& & & & \\
-0.015656249492 & (2,2) & &1& & & & & & & & & & & & & &\\
-0.012286435222 & (1,3) & & & & & &1& & & & & & & & & & \\
-0.012266183325 & (2,5) & & & & & & & & & & & & &1& & & \\
-0.007688713211 & (3,3) & & &1& & & & & & & & & & & & & \\
-0.005272945987 & (2,4) & & & & & & &1& & & & & & & & & \\
-0.003076941021 & (1,5) & & & & & & & & & &1& & & & & & \\
 \end{tabular}
 \end{center}
 \caption{Complete set of levels for the various sectors $(d,a)$ of the
   RSOS model, here for size $N=6$ and $p=6$ (the tricritical
   Potts model), cf.~Table~\ref{tab:levels}. The left column shows 
   $-N^{-1} \log \lambda$ (rounded to 12 digits), where $\lambda \ge 1$
   is the transfer matrix eigenvalue. The remaining columns show the
   corresponding multiplicities of $\lambda$ within the various sectors
   $T_L^{(\ast)}$ of the loop model (the asterisk $\ast$ denotes the
   blobbed sector, see Appendix A). A blank entry denotes zero multiplicity.
  The boundary weight $y=y(r)$
   is given by (\ref{yloop}), with $r$ indicated in the top of the table.}
 \label{tab:RSOS}
\end{table}

The first two cases coincide with what we discussed if $N$ is even and 
generalize it easily if $N$ is odd. In these two cases, the left 
boundary necessarily sees  
$b=d+1$, $c=d$, and the value of $y$ follows from the analysis of the 
loop model.
 
The last two cases meanwhile require that $b=d$, $c=d+1$. The new value 
of $y$ giving the correct weights to the boundary loops is thus 
$y=[d]_q/[d+1]_q$, which coincides with $y=[p+1-d]_q/[p-d]_q$.  

Note that the rules (\ref{RSOSrules}) are compatible with the global
symmetry $l_i \to p-l_i$ of the RSOS model configurations and weights
(\ref{Sl}). On the level of the sectors this reads
$(d,a) \to (p-d,p+1,a)$ and is nothing but the usual symmetry
of the Kac table (\ref{Kac}).

An extensive numerical check of (\ref{RSOSrules}) is given in
Table~\ref{tab:RSOS}. We show all levels observed for the RSOS model
with $N=6$ and $p=6$, along with the sector label $(d,a)$, and we give
the corresponding multiplicities of these levels in the loop model,
where each sector $T_L^{(\ast)}(r)$ is characterized by the number of
non contractible lines $L$, the blobbing of the leftmost string
($\ast$ denotes the blobbed sector), as well as $r$ which determines
the boundary weight $y(r)$ through (\ref{yloop}). Remarkably,
{\em all of the RSOS levels are also observed in the loop model}, with
the sector given precisely by the rules (\ref{RSOSrules}). This observation
extends to other values of $N$ and $p$ (with any parities).

Note that the dimension of the loop (resp.\ RSOS) model transfer matrix
does not (resp.\ does) depend on $r$. This apparent paradox is resolved
by the fact that a given loop model sector contains in general extra
eigenvalues (not shown in Table~\ref{tab:RSOS}) which are not present
in the RSOS model. It should also be noted that the dominant eigenvalue
in a given RSOS sector is always observed to be dominant as well in the
corresponding loop model sector (\ref{RSOSrules}).

Finally, we should mention a couple of fine details. In general the
levels are not observed to be degenerate. One exception visible in
Table~\ref{tab:RSOS} is that the level
$f=-0.036787385047$ is present in two distinct sectors, but since this
is so both on the RSOS and the loop side, definite labels respecting
(\ref{RSOSrules}) can be assigned as shown in the table. Another
exception is the level $f=-0.085859268861$ which apart from its
assignment to $T_4^\ast(r=1)$ by (\ref{RSOSrules}) appears also in two
more loop sectors, $T_2^\ast(r=1)$ and $T_4(r=1)$. This type of
degeneracies can be explained by quantum group arguments.


Recall now that the character of the irreducible representation of the 
Virasoro algebra with highest weight $h_{da}$ can be written as \cite{RochaCaridi}
\begin{equation}
\chi_{da}=\sum_{n=0}^\infty K_{d,a+2(p+1)n}-\sum_{n=1}^\infty K_{d,2n(p+1)-a}
\end{equation}
We wish to recover this character from the knowledge of the loop model 
partition function. This involves as usual an infinite series of 
additions and subtractions of sectors which is made possible in finite 
size by the quantum group symmetry. We suppose we are in the situation 
with heights $b=d,c=d+1$. To go from $d$ on the left side to $a$ on 
the right side we need to sandwich between the left and the right the 
adjacency matrix of the $A_{p}$ diagrams so that a random walk on this 
diagram, from boundary to boundary, hopping from non contractible 
cluster to non contractible cluster, takes one from $b$ to $a$. All 
the steps are identical with those in \cite{BauerSaleur} where 
partition functions with boundary conditions fixed at the leftmost and 
rightmost sides were computed. 
Using the eigenvectors of the adjacency matrix as in equation 
(4.15-4.19) of 
\cite{BauerSaleur}, the required 
expression is thus  (we have set $H=p+1$)
\begin{equation}
    {2\over 
    H}\sum_{p=1}^{H-1}\sin \left( {\pi pd\over H} \right) \sin \left( {\pi
      pa\over H} \right) 
    Z \left( \alpha={i\pi p\over H},\beta=d\alpha \right) \label{sumda}
    \end{equation}
Here $Z$ is calculated with $\alpha$ running over ${i\pi p\over H}$ and for 
each $\alpha$ the value of $m$---the weight of non contractible 
loops---follows from the same argument as in the case of the leading 
exponent $r=1$ in the bulk case, implying  $\beta=d\alpha$ so 
$m={\sinh(d+1)\alpha\over \sinh d\alpha}$, $\alpha={i\pi p\over H}$. Using 
the expression (\ref{niceZ})  we see that the functions $K_{d,d+2j}$ 
($r=d$) 
get a combinatorial factor
\begin{equation}
    {2\over H}\sum_{p=1}^{H-1}{1\over 2} \left[\cos(2j+d-a){p\pi\over 
    H}-\cos(2j+d+a){p\pi\over H}\right]
    \end{equation}
Since $d-a$ and $2j$ have the same parity, this select the conditions 
\begin{eqnarray}
    2j+d-a=2n_{1}H\nonumber\\
    2j+d+a=2n_{2}H
    \end{eqnarray}
 for the first and second   term respectively. Remembering that $j$ runs only from 
 $-[r/2]$ to infinity, and that when the foregoing conditions are 
 satisfied one gets a factor ${H\over 2}$ from the sum cancelling the 
 prefactor in (\ref{sumda}), it follows that
 \begin{equation}
     Z=\sum_{n=0}^\infty K_{d,a+2nH}-\sum_{n=1}^\infty K_{d,-a+2nH}
     \end{equation}
 which is the result we wanted.

\section{More algebraic considerations}
\label{sec:cabling}

There are many ways to think of the blob algebra. We would like 
now to think of it within the theory of cabling, i.e., tensor products of 
spin $1/2$ representations of $U_{q}(sl_{2})$. Consider therefore $r+1$ 
representations of spin $1/2$, and suppose we wish to project them on 
the maximally $q$-symmetric representation. The object doing this is 
the $q$-symmetrizer, whose expression is well known by induction to be
\cite{TLcabling}
 \begin{equation}
     {\cal S}_{r+1}(e_{1},\ldots,e_{r})={\cal S}_{r}- t_{r}{\cal S}_{r}
     e_{r}{\cal 
     S}_{r}
\end{equation}
with the boundary condition ${\cal S}_1 = 1$.
Here the $t_{r}$ are numbers given by 
 \begin{equation}
     t_{r}= {\sin r\gamma\over \sin (r+1)\gamma}
     \end{equation}
and the $e_{i}$ are TL generators which obey (\ref{TLalg}) as
before and act on the tensor product of the $i$'th and $(i+1)$'th 
spin $1/2$ representation. These generators will not be identified 
with the ones used in the previous sections however (see below).
The first values are well known: ${\cal S}_{1}=1$,
${\cal S}_{2}=1-{e_{1}\over 2\cos\gamma}$, etc. 
 
\begin{figure}
		    \begin{center}
			 \leavevmode
			 \epsfysize=60mm{\epsffile{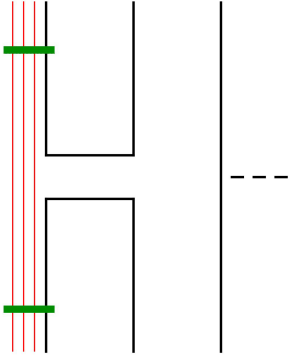}}
			 \end{center}
			 \protect\caption{(Color online.) Ghost strings $2-r,\ldots,0$,
         represented by thin red lines. They are symmetrized among themselves
         and with the first real line (the symmetrization is shown symbolically by horizontal green
         bars).}
			\label{ghost}
			\end{figure}

 Let us now add to our system $r-1$ ``ghost strings'' as in figure 
 \ref{ghost} (labelled 
 $2-r,\ldots,0$) on which 
 generators $e_{2-r},\ldots,e_{0}$ act, and let us symmetrize on 
 these ghost strings and the first one of our system. We then 
define  $b_{r}$  through
 \begin{equation}
     b_{r}={\cal S}_{r}(e_{2-r},\ldots,e_{0})
     \end{equation}
We have then
\begin{eqnarray}
    b_{1}&=&1\nonumber\\
    b_{2}&=&1-{e_{0 }\over 2\cos\gamma}\\
    b_{3}&=&1-t_{2}(e_{-1}+e_{0})+t_{1}t_{2}(e_{-1}e_{0}+e_{0}e_{-1})\nonumber
    \end{eqnarray}
and so on.
By construction, the $b_{r}$'s are projectors 
  \begin{equation}
      b_{r}^{2}=b_{r}\label{propi}
      \end{equation}
It is easy to show that they satisfy moreover
 \begin{equation}
     e_{1}b_{r}e_{1}={\sin (r+1)\gamma\over 
     \sin r\gamma}e_{1}P_{r-1}\label{propii}
     \end{equation}
where we denote by $P_{r-1}$ the symmetrizer on the ghost strings 
only, $P_{r-1}={\cal S}_{r-1}(e_{2-r},\ldots,e_{-1})$. Of course, 
$[e_{i},P_{r-1}]=0$ for all $i\geq 1$. We can thus consider a 
modified version of the TL algebra where instead of the generators 
$e_{i}$ we have the modified generators $\tilde{e}_{i}\equiv 
P_{r-1}e_{i}$. They obviously satisfy the bulk TL relations (\ref{TLalg}),
since $P_{r-1}$ is a projector. Moreover we now have the relations 
(\ref{blobalg}) of the blob algebra, with $b=b_r$ and (\ref{yloop}).
%
%
We have thus made the link with the foregoing discussion and shown 
that for $r$ an integer, the boundary conditions corresponding to the 
value (\ref{yloop}) can be obtained by adding ghost strings on the 
left boundary and symmetrizing them with the first ``real string'' in the 
system.

This  should not come as a surprise: it is easy to show 
using Pasquier's 6j calculations \cite{Etiology}
that the insertion of  $b_{r}$ through the ghost strings construction
translates into RSOS language by having heights increasing linearly 
from the left hand side of the ghost strings up to the first 
height from $l=1$ to $l=r$, and then because $b_{r}$ symmetrizes on 
$r$ strings and thus on the first real string as well, from $l=r$ to 
$l=r+1$. This proves that the transfer matrix $bT_{0}$ describes 
exactly the mixed RSOS boundary conditions. 

Note that one does not have to introduce the ghost strings. 
Algebraically, the same would be obtained by replacing these $r-1$ 
strings by a spin ${r-1\over 2}$ (in which case, $P_{r-1}=1$ 
identically). The operator $b_r$ then amounts to 
projecting the product of this representation and the first spin $1/2$ 
representation onto spin $r/2$. 
     
In this form,  the identification was already mentioned in 
\cite{BauerSaleur}. 

The formulas obtained in this section should match the ones  which 
appeared in a recent paper by Pearce et al. \cite{Pearce}; the 
derivation there is  based on boundary integrability, and does not 
refer to the blob or boundary Temperley Lieb algebra, as far as we can 
see.

\section{Conclusion}
\label{sec:conclusions}

In conclusion, it is important to stress that we have found a 
continuum of conformal boundary conditions for the dense  loop 
models. How to incorporate them into a consistent conformal field 
theory remains an open problem. Note that none of these boundary 
conditions involves the {\sl number of times loops touch the 
boundary}, which would correspond somehow to modified spin-spin 
couplings on the boundary. Rather, giving a different weight $y$ to 
loops touching the boundary can be interpreted in the $O(n)$ model 
most easily as restricting the 
degrees of freedom on the boundary to take values in a sub manifold of 
`dimension' $y$. This is not without reminding us of results in the 
WZW cases \cite{Alekseev}.  

\vskip.5cm

The results described in this paper point to many further directions. 
Among those are:

- geometric applications of the generating functions on the annulus

- derivation of the Bethe ansatz equations and of the spectrum of 
scaled gaps

- extensions to the dilute case

- extension to the double boundary case

- study of boundary conditions in $c=1$ theories

\noindent We hope to  report on these soon. To conclude, we now
give one example of application.
We consider the case 
$x=0$, $c=-2$ and the partition function with $x=l=0$, $y=m=1$ (i.e,. loops 
touching the boundary get a weight one, whether contractible or not, 
the others are not allowed). After some 
simple manipulations, (\ref{partfunc})  can be 
written as
\begin{equation}
    Z={q^{-1/24} \over P(q)}\sum_{j=-\infty}^{\infty}(-1)^{j}q^{(4j-1)^{2}/32}
    \end{equation}
This is  easy to interpret geometrically as the partition function of a gas 
of dense loops, all contractible, which are all constrained to touch 
the left boundary. The loops can also be replaced by trees. Meanwhile, 
this partition function can also be interpreted in the symplectic
fermion theory \cite{Volkerunpublished}.

\vskip.5cm
\noindent{\bf  Acknowledgments:} H. Saleur thanks I. Kostov and V. 
Schomerus for many interesting discussions, and for communication of 
their unpublished results \cite{Kostovprivcomm,Volkerunpublished}.

\vskip.5cm
\noindent{\bf Note added in proof:} 
After the completion of this work, I. Kostov has studied (in the preprint
hep-th/0703221) the coupling of our boundary loop model to two-dimensional
quantum gravity. His results corroborate those presented here.

\appendix

\section{Transfer matrix structure}

In this Appendix we discuss the construction and structure of the
transfer matrix of the boundary loop model, corresponding to taking
the limit $\lambda\to\infty$ in (\ref{TMlambda}).

We recall the most general case of the boundary loop model,
cf.~Fig.~\ref{partfunct}. The model is defined on an annulus with the
exterior boundary being distinguished. Each loop touching at least once this
boundary gets a weight $y$ if it is contractible (i.e., homotopic to a point), and $m$ if it is
not. Each loop that never touches the distinguished boundary gets
a weight $x$ if it is contractible, and $l$ if it is not.
For an annulus of width $N$ strands (we assume $N$ even) and circumference $M$, this defines a
partition function $Z_{N,M}(x,y,l,m)$ which can be expressed in terms
of the $M$'th power of the transfer matrix $T=b T_0$, where $T$ is given in terms of TL generators
by (\ref{TM}),  and the generators $b$, $e_i$ obey the blob algebra $B_N(x,y)$ defined by (\ref{TLalg})--(\ref{blobalg}).

\begin{figure}
  \begin{center}
    \leavevmode
    \epsfysize=80mm{\epsffile{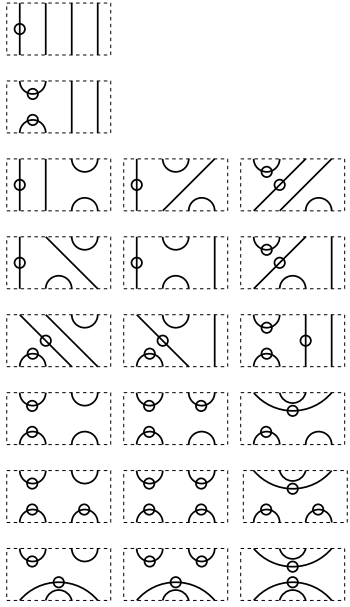}}
  \end{center}
  \protect\caption{Ordering of the states of the transfer matrix $T$, here for width
    $N=4$ strands.}
  \label{app:states}
\end{figure}

For simplicity we henceforth represent the annulus as a rectangle of width $N$
and height $M$, with periodic boundary conditions identifying its top and
bottom sides.  The distinguished boundary is taken to be the left side. The transfer matrix then acts on states which can be depicted
graphically as fully-packed non-crossing link patterns within a slab bordered by two horizontal
rows, each of $N$ points. A link joining the top and the bottom row is called
a {\em string}, and any other link is called an {\em arc}. The action of a word in $\mathcal{B}_N(x,y)$
on a state is obtained by adjoining the word to the top row of
the slab (i.e., time propagates upwards).  The 20 possible states for $N=4$
are represented in Fig.~\ref{app:states}.

Note that links touching a point on
the left boundary are necessarily blobbed (shown by a circle in Fig.~\ref{app:states}).
Only links up to and including the leftmost string can be blobbed. The states
can be ordered as follows: First we sort the states according to a decreasing number of strings $L$. For fixed
$L>0$, we place first the states in which the leftmost string is unblobbed.
And finally we group together states with fixed $L$ and fixed blobbing of the
leftmost string, according to the link/arc configuration of the lower row of
the slab. This gives the order of the rows of Fig.~\ref{app:states}. The
ordering of states within each row is according to the link/arc configuration
of the upper row of the slab.

With this ordering of the states, $T$ has a blockwise lower triagonal
structure,  with each block corresponding to a group of states as defined
above. The reason is that acting by $e_i$ can annihilate two strings (if their
positions on the top side of the slab are $i$ and $i+1$) but cannot create any
strings. Likewise, acting by $b$ can blob the leftmost string, but it cannot
subsequently be unblobbed.  The triagonal structure implies that the eigenvalues of $T$ are the union of eigenvalues of the
blocks on its diagonal. Moreover, blocks differing only by the configuration
of the bottom  row are identical. For the purpose of studying only the spectrum
of $T$ the bottom row can therefore be completely forgotten, leading to a much
smaller transfer matrix.

\begin{figure}
  \begin{center}
    \leavevmode
    \epsfysize=25mm{\epsffile{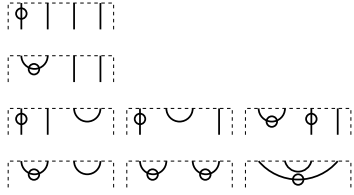}}
  \end{center}
  \protect\caption{Reduced states for $N=4$.}
  \label{app:red_states}
\end{figure}

Indeed let us define a {\em reduced state} as a non-crossing link
pattern on $N$ points. A (full) state can be turned into a pair of
reduced states by cutting each of its strings and pulling apart the
upper and lower parts. For convenience, a cut string will still be
called a string with respect to the reduced state. For $N=4$ there are
8 reduced states, shown in Fig.~\ref{app:red_states}.

The blocks on the diagonal of $T$ are denoted $T_L$ and $T_L^\ast$,
where $L$ is the number of strings, and the presence (resp.\ absence)
of the asterisk ($\ast$) indicates that the leftmost link is blobbed
(resp.\ unblobbed).  Note that the blocks $T_L$ and $T_L^\ast$ can be
constructed in terms of the reduced states.  The numerical studies of
the spectral properties of $T$ reported in this paper were done by
diagonalizing these blocks in the basis of reduced states.
Their dimensions read (see Appendix B for proofs of closely related statements)
\begin{eqnarray}
 {\rm dim}\ T_L &=& {N-1 \choose (N-L-2)/2}
 \ \ \mbox{ for $L=0,2,\ldots,N-2$}
 \nonumber \\
 {\rm dim}\ T_L^\ast &=& {N-1 \choose (N-L)/2}
 \ \ \ \ \  \ \ \mbox{ for $L=2,4,\ldots,N$}
\end{eqnarray}

With the terminology being fixed, the annulus partition function can
now be written as \cite{SalasRing}
\begin{equation}
 Z_{N,M}(x,y,l,m) = \left \langle u | T^M | v \right \rangle \,.
\label{app:partfunc}
\end{equation}
Here, the right vector $| v \rangle$ is the unit vector corresponding
to the state with $N$ strings. The left vector $\langle u |$ is obtained
by identifying the top and bottom rows for each state; counting the number of 
loops of each type (contractible or not, blobbed or not) gives the
corresponding weight as a monomial in $x$, $y$, $l$, and $m$. For instance,
with $N=4$ and the ordering of the states shown in Fig.~\ref{app:states},
we have
\begin{eqnarray}
 |u\rangle &=& (1,0,0,\ldots,0) \nonumber \\
 |v\rangle &=& (l^3 m, l^2 y, l m x, l m, y,\ldots, x y)
\end{eqnarray}

\section{Exact eigenvalue amplitudes}

The goal of this Appendix is to provide a rigorous combinatorial proof
of the amplitude formulae (\ref{guess}) by generalizing the working of
\cite{RichardCyclic}. The discussion assumes knowledge of the transfer
matrix blocks $T_L$ and $T^\ast_L$ defined in Appendix A.

Following \cite{RichardCyclic}, we introduce the {\em characters}
\begin{equation}
 K_k = {\rm Tr}\, (T_{2k})^M, \qquad
 K_k^\ast = {\rm Tr}\, (T_{2k}^\ast)^M \,,
 \label{app:charK}
\end{equation}
where we stress that the trace is over {\em reduced} states.
Also, let $Z_j$ (resp.\ $Z_j^\ast$) be the annulus partition function constrained to have exactly
$2j$ unblobbed non contractible loops (resp.\ $2j-1$ unblobbed and $1$ blobbed
non contractible loops). In other words, $Z_j$ (resp.\ $Z_j^\ast$) consists of
the terms in the full partition function $Z_{N,M}(x,y,l,m)$ whose $l,m$
dependence is $l^{2j}$ (resp.\ $l^{2j-1} m$).
The goal is to search for a decomposition of the form
\begin{eqnarray}
 Z_j &=& \sum_{k=j}^{N/2} \left[ D^{(j)}_k l^{2k} K_k + D^{(j)}_{k\ast}
   l^{2k-1} m K_k^\ast \right] \, \nonumber \\
 Z_j^\ast &=& \sum_{k=j}^{N/2} \left[ D^{(j)\ast}_k l^{2k} K_k + D^{(j)\ast}_{k\ast}
   l^{2k-1} m K_k^\ast \right] \,,
\label{app:decomp1}
\end{eqnarray}
where $D^{(j)}_k$, $D^{(j)}_{k\ast}$, $D^{(j)\ast}_k$, and
$D^{(j)\ast}_{k\ast}$ are coefficients to be determined. In the notation of
(\ref{guess}) we have then
\begin{eqnarray}
 D_{2j}^\ast &=& \sum_{i=0}^j \left[ D^{(i)}_{j\ast} l^{2i} +  D^{(i)\ast}_{j\ast} l^{2i-1} m
 \right] \nonumber \\
 D_{2j} &=& \sum_{i=0}^j \left[ D^{(i)}_j l^{2i} + D^{(i)\ast}_{j} l^{2i-1} m
\right] 
\label{app:decomp2}
\end{eqnarray}
with $D^{(j)}_{0\ast} = D^{(j)\ast}_{0\ast} = D^{(j)\ast}_{j} = 0$.

Rather than solving directly for the decomposition of $Z_j$ in terms of $K_k$,
the idea \cite{RichardCyclic} is now to turn the problem upside down and
look for the decomposition of $K_k$ in terms of $Z_j$:
\begin{eqnarray}
 K_k &=& \sum_{j=k}^{N/2-1} E^{(k)}_j \frac{Z_j}{l^{2j}} + \sum_{j=k+1}^{N/2} E^{(k)}_{j\ast}
 \frac{Z_j^\ast}{l^{2j-1} m} \nonumber \\
K_k^\ast &=& \sum_{j=k}^{N/2-1} E^{(k)\ast}_j \frac{Z_j}{l^{2j}} + \sum_{j=k}^{N/2}
E^{(k)\ast}_{j\ast} \frac{Z_j^\ast}{l^{2j-1} m} \,.
\end{eqnarray}

\begin{figure}
  \begin{center}
    \leavevmode
    \epsfysize=60mm{\epsffile{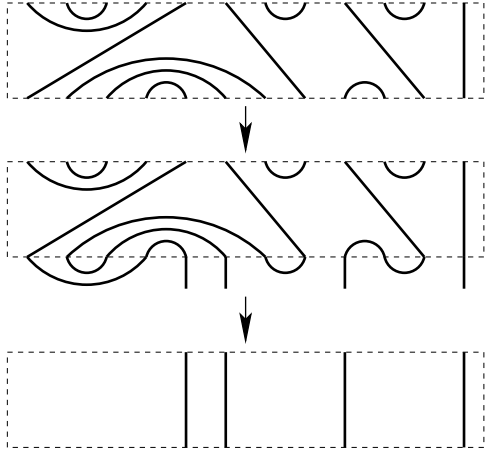}}
  \end{center}
  \protect\caption{Construction of invariant restricted states.
    (a) A configuration contributing to $Z_j$ with $N=12$ and $j=2$, here
    depicted as a state. (b) Application on the bottom of the reduced
    state corresponding to the top half of (a). (c) After removal of
    arcs one has simply $2j$ links.}
  \label{app:compstate}
\end{figure}

The determination of the coefficients $E$ can be turned into a combinatorial
counting problem as follows. First, recall that $K_k$ and $K_k^\ast$ were
defined as {\em traces} over restricted states (in contradistinction to the
partition function which, as we have seen in (\ref{app:partfunc}), is more
complicated than just a trace).  We must now determine how many times each $Z_j$ and $Z_j^\ast$
occurs within a given trace. Consider therefore some configuration ${\cal C}$ on the
annulus that contributes to (say) $Z_j$ (i.e., has $2j$ non contractible unblobbed
loops). An example with $j=2$ and $N=12$ is shown in Fig.~\ref{app:compstate}a.
It is convenient to not represent the contractible loops within the
configuration, i.e., to depict it as a state. This configuration will
contribute to the trace only over
such restricted states ${\cal S}$ that are left {\em invariant} by the action of the configuration.
Therefore, ${\cal S}$ must contain the same arcs as does ${\cal C}$ in its top
row (see Fig.~\ref{app:compstate}b). It suffices therefore to determine the
parts of ${\cal S}$ which connect onto the starting points of the $2j$ non
contractible loops (see Fig.~\ref{app:compstate}c). Since the goal is to
determine the contribution to (say) $K_k$, precisely $2k$ strings and $j-k$
arcs must be used.

With this in mind, the coefficients $E$ are then determined as the following
counting problems. In all cases, construct a reduced state on $2j$ strands,
using $2k$ strings and $j-k$ arcs. Further:
\begin{itemize}
 \item For $E_j^{(k)}$, all strings are unblobbed, but the {\em exterior} arcs to the left of
   the first string may be blobbed.
 \item For $E_j^{(k)\ast}$, the problem is the same, except that the leftmost
  string must be blobbed. But evidently this leads to the same counting, and
  so $ E^{(k)\ast}_j = E^{(k)}_j$.
 \item For $E_{j\ast}^{(k)}$, the leftmost strand becomes blobbed (since we are considering
  a contribution from $Z_j^\ast$), and so must connect onto a blobbed object
  (arc or string). But as the strings are unblobbed (since we are considering a
  contribution from $K_k$), it follows that the leftmost object is a blobbed
  arc.
 \item For $E_{j\ast}^{(k)\ast}$, the leftmost object (arc or string) as well
   as the leftmost string must be blobbed.
\end{itemize}

These counting problems are easily solved using generating function
techniques. As a warmup, consider the counting of restricted states made up of
only arcs. Associate to each {\em pair} of sites an activity $z$. A state is
either empty, or has a leftmost arc which divides the space into two parts
(inside the arc and to its right) each of which can accommodate an independent arc state. The generating
function $f(z)$ therefore satisfies $f(z) = 1 + z [ f(z) ]^2$ with regular
solution
\begin{equation}
 f(z) = \frac{1 - \sqrt{1-4z}}{2z} = \sum_{n=0}^\infty C_n
 \, z^n\,.
\end{equation}
The coefficients are the celebrated Catalan numbers $C_n = \frac{(2n)!}{n!(n+1)!}$.

Consider next states made up of only arcs, but in which exterior arcs may (but
need not) be blobbed. Call the generating function $g(z)$.  If the state is non-empty, the leftmost arc is
necessarily exterior. Inside it are $f(z)$ states, and to its right $g(z)$
states. Thus, $g(z) = 1+2z f(z) g(z)$, or
\begin{equation}
 g(z) = \frac{1}{\sqrt{1-4z}} = \sum_{n=0}^\infty {2n \choose n} z^n \,.
\end{equation}

We can now attack the case of $E_j^{(k)}$. Since there are $2k$ strings (all
unblobbed), all of which divide the space into independent parts, the
generating function reads
\begin{equation}
 h_k(z) = g(z) \times z^k \left[ f(z) \right]^{2k} = \sum_{n=k}^\infty {2n
   \choose n-k} z^n
\end{equation}
and we infer that $E_j^{(k)} =  E^{(k)\ast}_j = {2j \choose j-k}$.

For $E_{j\ast}^{(k)}$ it follows from the above observations that the
generating function with $2k$ strings reads
\begin{equation}
 i_k(z) = z f(z) h_k(z) = \sum_{n=k+1}^\infty {2n-1 \choose n-k-1} z^n
\end{equation}
and so $E_{j\ast}^{(k)} = {2j-1 \choose j-k-1}$.

Finally, for $E_{j\ast}^{(k)\ast}$ we must distinguish between the cases where the
leftmost object is an arc or a string. This gives the generating function
\begin{equation}
 j_k(z) = \left(1 + z f(z) g(z) \right) z^k \left[ f(z) \right]^{2k}
            = \sum_{n=k}^\infty {2n-1 \choose n-k} z^n
\end{equation}
whence $E_{j\ast}^{(k)\ast} = {2j-1 \choose j-k}$ .
 
To summarize, we have shown that
\begin{eqnarray}
 K_k &=& \sum_{j=k}^{N/2-1} {2j \choose j-k} \frac{Z_j}{l^{2j}} +
 \sum_{j=k+1}^{N/2} {2j-1 \choose j-k-1}
 \frac{Z_j^\ast}{l^{2j-1} m} \nonumber \\
K_k^\ast &=& \sum_{j=k}^{N/2-1} {2j \choose j-k}  \frac{Z_j}{l^{2j}} + \sum_{j=k}^{N/2}
{2j-1 \choose j-k}\frac{Z_j^\ast}{l^{2j-1} m} \,.
\end{eqnarray}
It is easily checked that this system of equations is invertible, and after
some straightforward manipulations the $Z_j$ and $Z_j^\ast$ can be isolated.
In the notation of (\ref{app:decomp1}) the solution reads
\begin{eqnarray}
 D_k^{(j)} = (-1)^{j+k} {j+k \choose 2k}, \qquad
 D_{k \ast}^{(j)} = (-1)^{j+k-1} {j+k-1 \choose 2k-1} \nonumber \\
 D_k^{(j)\ast} = (-1)^{j+k} {j+k-1 \choose 2k}, \qquad
 D_{k \ast}^{(j)\ast} = (-1)^{j+k} {j+k-1 \choose 2k-1} 
\end{eqnarray}
Applying (\ref{app:decomp2}) then finally leads to (\ref{guess}).

\end{document}